\documentclass[fleqn,usenatbib]{mnras}

\usepackage{mathptmx}
\usepackage{txfonts}

\usepackage[T1]{fontenc}
\usepackage{ae,aecompl}
\usepackage{lipsum}
\usepackage{soul}

\usepackage{graphicx}	
\usepackage{amssymb}	

\usepackage{color} 

\def\ergcms{{\rm erg\,cm^{-2}\,s^{-1}}}

\def\Kbb{{K_{\rm bb}}}
\def\Rbb{{R_{\rm bb}}}
\def\Tbb{{T_{\rm bb}}}

\def\Kdbb{{K_{\rm dbb}}}

\def\Tdbb{{T_{\rm dbb}}}
\def\Fdbb{{F_{\rm dbb}}}

\def\Mns{{M_{\rm NS}}}
\def\Rns{{R_{\rm NS}}}

\def\rxte{\textit{RXTE}}

\def\ergcms{{\rm erg\,cm^{-2}\,s^{-1}}}
\def\cm2{{\rm cm^{-2}}}

\def\Te{T_{\rm e}}
\def\NH{N_{\rm H}}

\renewcommand{\deg}{\ensuremath{^{\circ}}}
 
\title[Spreading layer in 4U 1608--52 during X-ray bursts]{Variable spreading layer in 4U 1608--52 during thermonuclear X-ray bursts in the soft state}

\author[J.~J.~E. Kajava et al.]{J.~J.~E. Kajava,$^{1,2}$ K.~I.~I. Koljonen,$^{1,3}$ J. N\"{a}ttil\"{a},$^{2,4}$  V. Suleimanov$^{5,6}$ and \newauthor J. Poutanen$^{2,4,7}$
\\
$^{1}$Finnish Centre for Astronomy with ESO (FINCA), University of Turku, V\"{a}is\"{a}l\"{a}ntie 20, FI-21500 Piikki\"{o}, Finland
 \\
$^{2}$Tuorla Observatory, Department of Physics and Astronomy, University of Turku, V\"{a}is\"{a}l\"{a}ntie 20, FI-21500 Piikki\"{o}, Finland\\
$^{3}$Aalto University, Mets\"{a}hovi Radio Observatory, Mets\"{a}hovintie 114, FI-02540 Kylm\"{a}l\"{a}, Finland\\
$^{4}$Nordita, KTH Royal Institute of Technology and Stockholm University, Roslagstullsbacken 23, SE-10691 Stockholm, Sweden \\
$^{5}$Institut f\"ur Astronomie und Astrophysik, Kepler Center for Astro and Particle Physics, Universit\"at T\"ubingen, Sand 1, \\D-72076 T\"ubingen, Germany \\
$^{6}$Kazan (Volga region) Federal University,  Kremlevskaya str. 18, Kazan 420008, Russia\\
$^{7}$Kavli Institute for Theoretical Physics, University of California, Santa Barbara, CA 93106, USA 
}

\date{Accepted 2017 July 28. Received 2017 July 28; in original form 2017 May 10}

\pubyear{2017}

\begin{document}
\label{firstpage}
\pagerange{\pageref{firstpage}--\pageref{lastpage}}
\maketitle

\begin{abstract}
Thermonuclear (type-I) X-ray bursts, observed from neutron star (NS) low-mass X-ray binaries (LMXB), provide constraints on NS masses and radii and consequently the equation of state of NS cores.
In such analyses various assumptions are made without knowing if they are justified.
We have analyzed X-ray burst spectra from the LMXB 4U 1608--52, with the aim of studying how the different persistent emission components react to the bursts.
During some bursts in the soft spectral state we find that there are two variable components; one corresponding to the burst blackbody component and another optically thick Comptonized component.
We interpret the latter as the spreading layer between the NS surface and the accretion disc, which is not present during the hard state bursts.
We propose that the spectral changes during the soft state bursts are driven by the spreading layer that could cover almost the entire NS in the brightest phases due to the enhanced radiation pressure support provided by the burst, and that the layer subsequently returns to its original state during the burst decay.
When deriving the NS mass and radius using the soft state bursts two assumptions are therefore not met: the NS is not entirely visible and the burst emission is reprocessed in the spreading layer, causing distortions of the emitted spectrum.
For these reasons the NS mass and radius constraints using the soft state bursts are different compared to the ones derived using the hard state bursts. 
\end{abstract}


\section{Introduction}

Thermonuclear (type I) X-ray bursts are triggered by unstable hydrogen and helium burning in a neutron star (NS) envelope (see \citealt{LvPT93}, for review).
The fuel is provided by a low-mass binary companion, from which gas is accumulated onto the NS through an accretion disc.
The thermonuclear explosions can occasionally be so intense that the Eddington limit is exceeded for a few seconds, which causes the entire NS photosphere to expand and subsequently contract back towards the NS surface \citep{SBB77,HCL80,Paczynski83}.
These photospheric radius expansion (PRE) bursts form a very interesting subset of X-ray bursts, because they allow the determination of the NS radius $\Rns$ and mass $\Mns$ simultaneously (see \citealt{NSK16, OPG16, SPN17}, for most recent measurements).
Consequently, PRE-bursts can be used to place constraints on the NS equation of state models that describe the properties of the ultra dense matter in NS cores \citep{Lattimer12}. 

In low-mass X-ray binaries (LMXB) the accretion onto the NS may proceed in two distinct ways, which manifest themselves as spectral states.
Broadly speaking, at the lowest fluxes LMXBs are in the so called hard spectral state (or island state), whereas at higher fluxes LMXBs are in the soft spectral state (see the review by \citealt{DGK07} for NS-LMXB spectral state classifications).
In many respects the spectral hysteresis patterns are similar to the ones seen in the black hole binaries (see, e.g., \citealt{MDFM14}), but the NS surface plays a crucial role in creating differences. 
In the hard state, when the accretion disc near the NS is thought to puff up to an optically thin and geometrically thick flow, the hot electrons, protons and ions deposit most of their energy at the upper atmosphere layers \citep{BSW92}.
This heated NS atmosphere then acts as an extra source of soft seed photons for Comptonization in the hot flow, leading to cooler equilibrium electron temperatures and softer X-ray spectra than in black hole binaries \citep[e.g.][]{DGK07,BGS17}. 
Only in the soft state the hot inner flow may collapse into a thin disc that extends all the way down to the NS surface.
In this case an optically thick boundary/spreading layer forms in between the disc and the NS surface \citep{IS99,IS10,SP06}, which emits slightly hotter quasi-thermal emission than the accretion disc with roughly the same luminosity \citep{GRM03,RG06,RSP13}.

In order to measure $\Rns$ and $\Mns$ using X-ray bursting NS-LMXBs, several of the following simplifying assumptions are commonly made:
\begin{enumerate}
\item the NS is entirely visible during the X-ray burst, rather than being partially obscured by the accretion disc and the spreading layer \citep{LS85}; 
\item the burst emission is isotropic, although the reprocessing in the disc can cause the emission to be anisotropic \citep{LS85}; 
\item  X-ray burst spectra are well described by a blackbody, but clear deviations are known to occur (e.g., \citealt{WGP15});
\item  the emission region radius  can be obtained from the observed blackbody radii using colour corrections from the NS atmosphere models \citep[e.g.][]{SPW12}, but the majority of bursts  are inconsistent with these model predictions, particularly those occurring in the soft spectral state \citep{KNL14};
\item  the NS photospheric composition remains constant during the bursts, while there is evidence that the nuclear burning ashes produced in the bursts may reach the photosphere, changing the spectral hardening (i.e. colour correction factors, \citealt{NSK15}) and imprinting photo-ionization edges to the spectra \citep{inZW10, KNP17}, and 
\item  the persistent (accretion) emission stays constant during the burst, even though recently several findings suggest that the persistent emission components are disturbed by the X-ray bursts (e.g., \citealt{WGP15}).
\end{enumerate}
It is possible that all, some or even none of these assumptions are actually reasonable, and it remains contentious to what extent they can affect the NS mass and radius constraints.

In this paper, we study the above effects  in the X-ray bursts observed from the transient NS-LMXB 4U 1608--52.
In Section 2, we describe how we have reduced and analyzed the data used in this study. 
In Section 3, we mainly focus on the finding that one of the spectral components detected in the persistent emission is also strongly variable during the soft state X-ray bursts.
This component is consistent with the optically thick spreading layer, which is not found in the hard state bursts.
The behavior in the soft state bursts of 4U 1608--52 is similar to the spectral evolution seen during to the superburst of 4U 1636--536 \citep{KKK16}, whereas the X-ray bursts taking place during the hard spectral state are much better described by the blackbody model.
In Section 4, we discuss the main implication of our results, namely, that the spreading layer may occasionally cover almost the entire NS. 
This effect is not accounted for in studies that use soft state X-ray bursts for NS mass and radius determination (e.g. \citealt{GOC10, OPG16}), and likely causes the differences in NS mass and radius constraints obtained from the hard state bursts \citep{PNK14}. 

\section{RXTE/PCA observations of 4U 1608--52}

We have identified and extracted time resolved spectra of all X-ray bursts from 4U 1608--52 that have been observed with the Proportional Counter Array (PCA) instrument  \citep{JMR06} onboard the \textit{Rossi X-ray Timing Explorer} (\rxte) spacecraft.
We have also extracted X-ray spectra of the accretion (persistent) spectrum outside the burst intervals, in order to characterize the spectral states of this LMXB. 
These data have already been analyzed and presented in \citet{PNK14} and \citet{KNL14} and, therefore, we only repeat here the most relevant points.

The PCA burst data were extracted using 0.25--4~s time intervals depending on the instrument count rate.
We employed two different methods to model these spectra.
The first method used the standard approach -- used also in \citet{PNK14} and \citet{KNL14} -- where the accretion spectra prior to X-ray bursts were subtracted as the background.
In the second method we extracted a 160~s segment prior to the burst from the ``standard 2'' PCA data mode (using only the events recorded by the top layer of PCU2), modeled it as described below, and then fixed some model parameters to the best-fitting values when modeling the burst spectra.
We used shorter exposure times for bursts that took place less than 160~s from the start of a stable spacecraft pointing.

The spectra were deadtime corrected, and they were fitted in the 3--20 keV spectral range in \textsc{xspec} v.12.8 \citep{Arnaud1996}.
In all cases the interstellar absorption was modeled using the \textsc{tbabs} model \citep{WAM00}, with the hydrogen column density fixed to $\NH = 0.89\times 10^{22}\, \cm2$ \citep{KintZK08}.
The burst emission was described with the blackbody model, yielding a blackbody temperature $\Tbb$ and normalization $\Kbb=(\Rbb/d_{10})^2$, where $\Rbb$ is the blackbody radius in units of km and $d_{10}$ is the distance to the source in units of 10~kpc.
The persistent emission was modeled using three components: the accretion disc (\textsc{diskbb}), the spreading layer (\textsc{comptt}) and a weak iron emission line (\textsc{gauss}) that was present in most of the persistent emission data sets. 
The fluxes were obtained using the \textsc{cflux} convolution model and the model errors were derived with the \textsc{error} command at 1$\sigma$ confidence level.

\section{Results}

\subsection{NMF spectral decomposition of bursts}

\begin{figure}
\centering
\includegraphics[width=\linewidth]{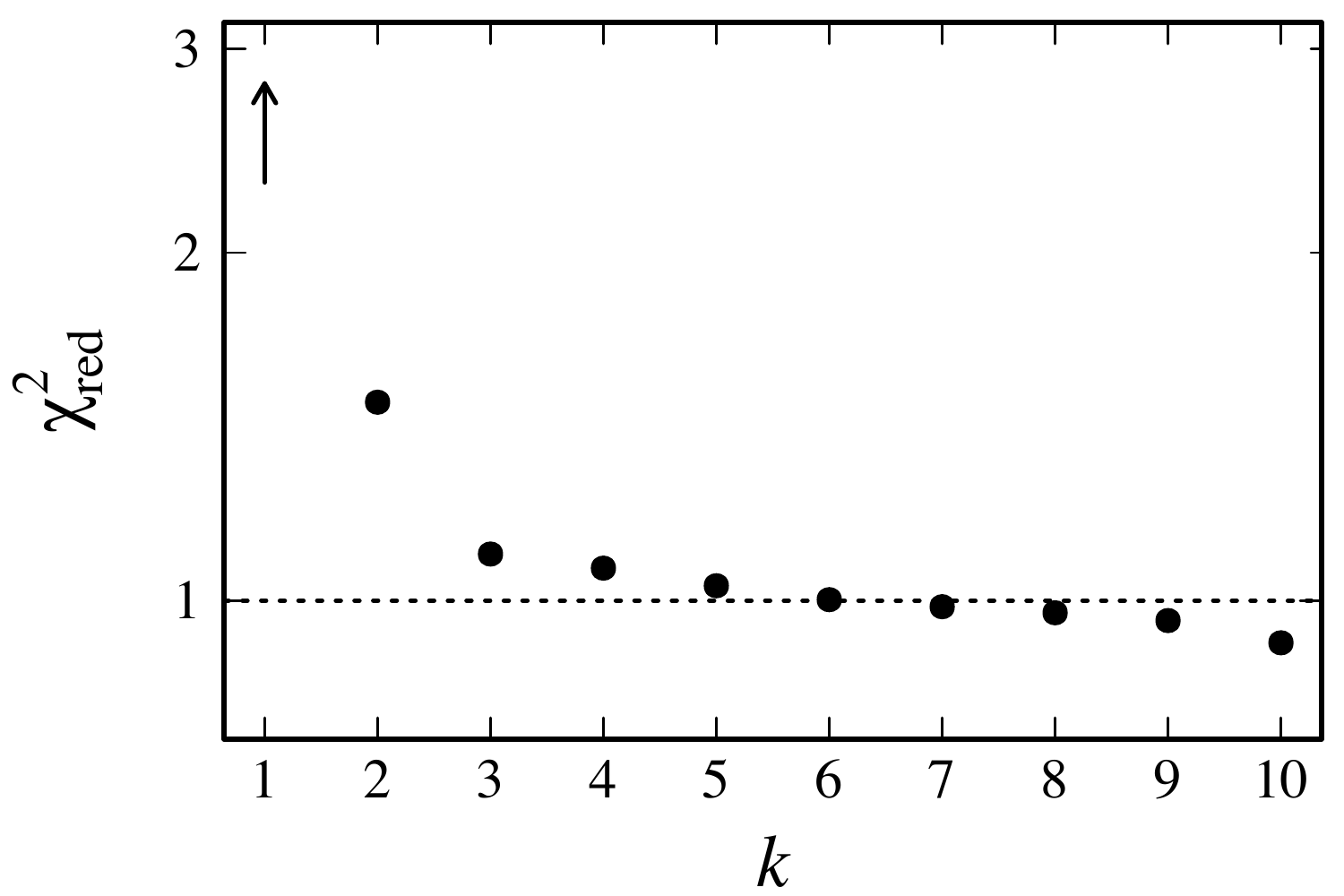}
\caption{The $\chi^{2}$-diagram used for the determination of the degree of factorization for NMF analysis using data from the four soft-state PRE-bursts of 4U 1608--52 (not subtracting the background) that have the most non-Planckian spectra (see Fig.~\ref{fig:softstatebursts}). The diagram shows a kink at $k=3$ after which further increase of the degree of factorization reduces the $\chi^{2}_{\rm red}$-value only insignificantly. Similarly to the 4U 1636--536 superburst \citep{KKK16}, three components are enough to explain the spectral variability during the soft state PRE bursts of 4U 1608--52. The $\chi^{2}_{\rm red}$-value for the one-component factorization is $\approx$28. 
}
\label{fig:nmfchi}
\end{figure}

After re-inspecting the blackbody fitting results for the soft state PRE bursts presented in \citet{PNK14} and \citet{KNL14}, we identified four PRE bursts that  have  large  reduced $\chi^2$ values (up to $\sim 4$), i.e. they showed clear deviations from the Planckian spectra.
These are the bursts observed in 1998 March 27 (OBSID 30062-01-01-00), 2002 September 09 (OBSID 70059-01-21-00), 2002 September 12 (OBSID 70059-03-01-000) and 2011 June 13 (OBSID 96423-01-11-01).
Following \citet{KKK16}, we used the non-parametric non-negative matrix factorization (NMF) spectral decomposition technique to determine the possible causes behind the poor fits.
In NMF, the observations $X_{ji}$, where $i$ is the observation number and $j$ is the spectral channel number, are approximated as a linear decomposition of a small number ($k$) of NMF (spectral) components $W_{jk}$ that vary in normalization according to their signal $S_{ki}$, i.e. $X_{ji} \approx \sum_{k} W_{jk} S_{ki}$. $W$ and $S$ are found iteratively by minimizing a cost function (generalized Kullback-Leibler divergence) and requiring that $W$ and $S$ retain only positive values. We refer the reader to \citet{Koljonen2015}, and references therein for a more detailed discussion about the NMF.  

We first determined the degree of NMF factorization using the $\chi^2$-diagram method \citep{Koljonen2015, KKK16}, shown in Fig. \ref{fig:nmfchi}.
We find that, similarly to the 4U 1636--536 superburst \citep{KKK16}, three NMF components are needed to describe the spectral variability during the bursts.

\begin{figure}
\centering
\includegraphics[width=\linewidth]{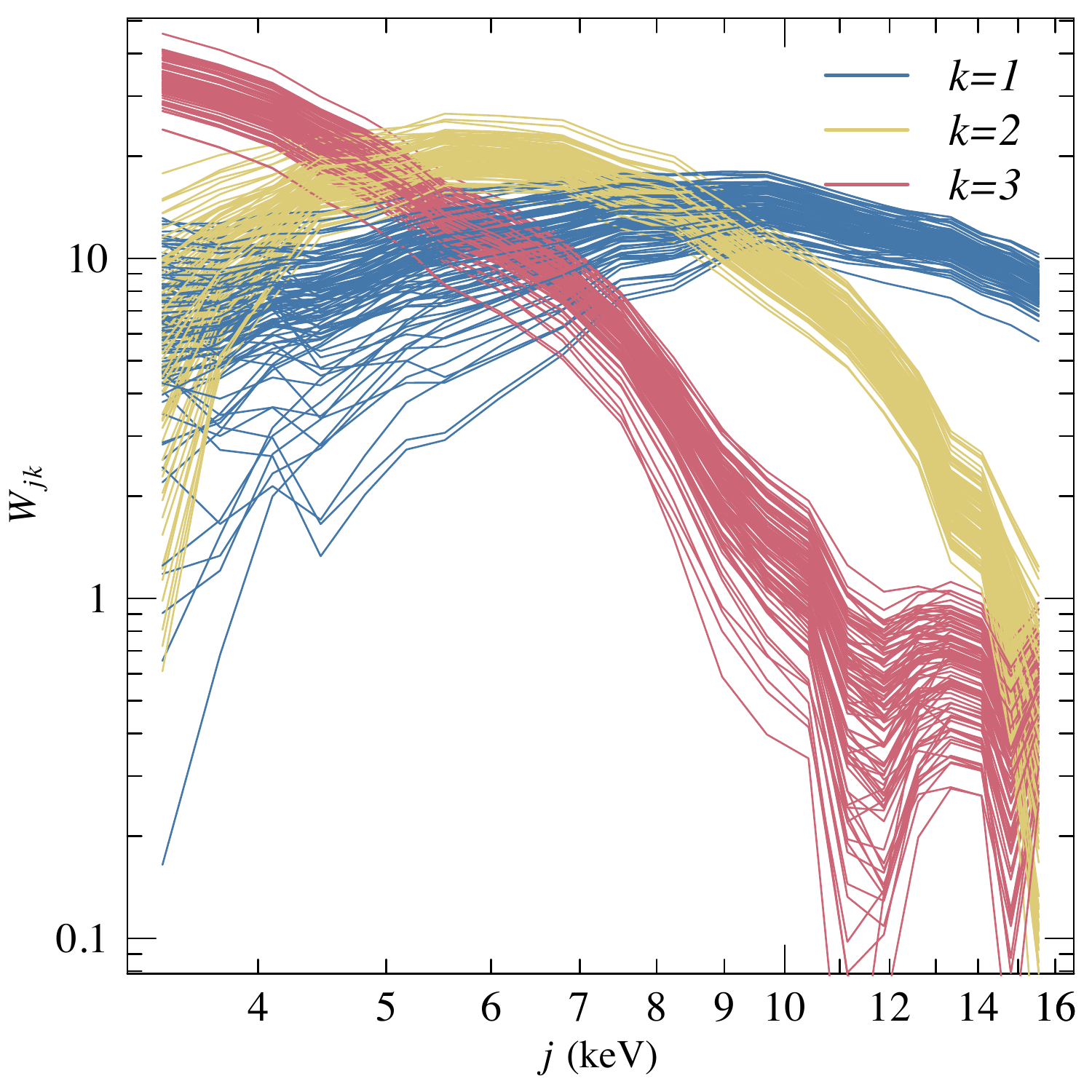}
\caption{A sample of spectra ($W_{jk}$) from multiple NMF runs on the four non-Planckian soft state PRE bursts of 4U 1608--52 (background was not subtracted). The spectra have the same overall shape as in the 4U 1636--536 superburst data; the $k=1$ thus corresponds to the constant temperature spreading layer component and the $k=2$ and $k=3$ components together can be interpreted as the variable temperature burst emission.
}
\label{fig:nmfcomponents}
\end{figure}

\begin{figure*}
\centering
\includegraphics[]{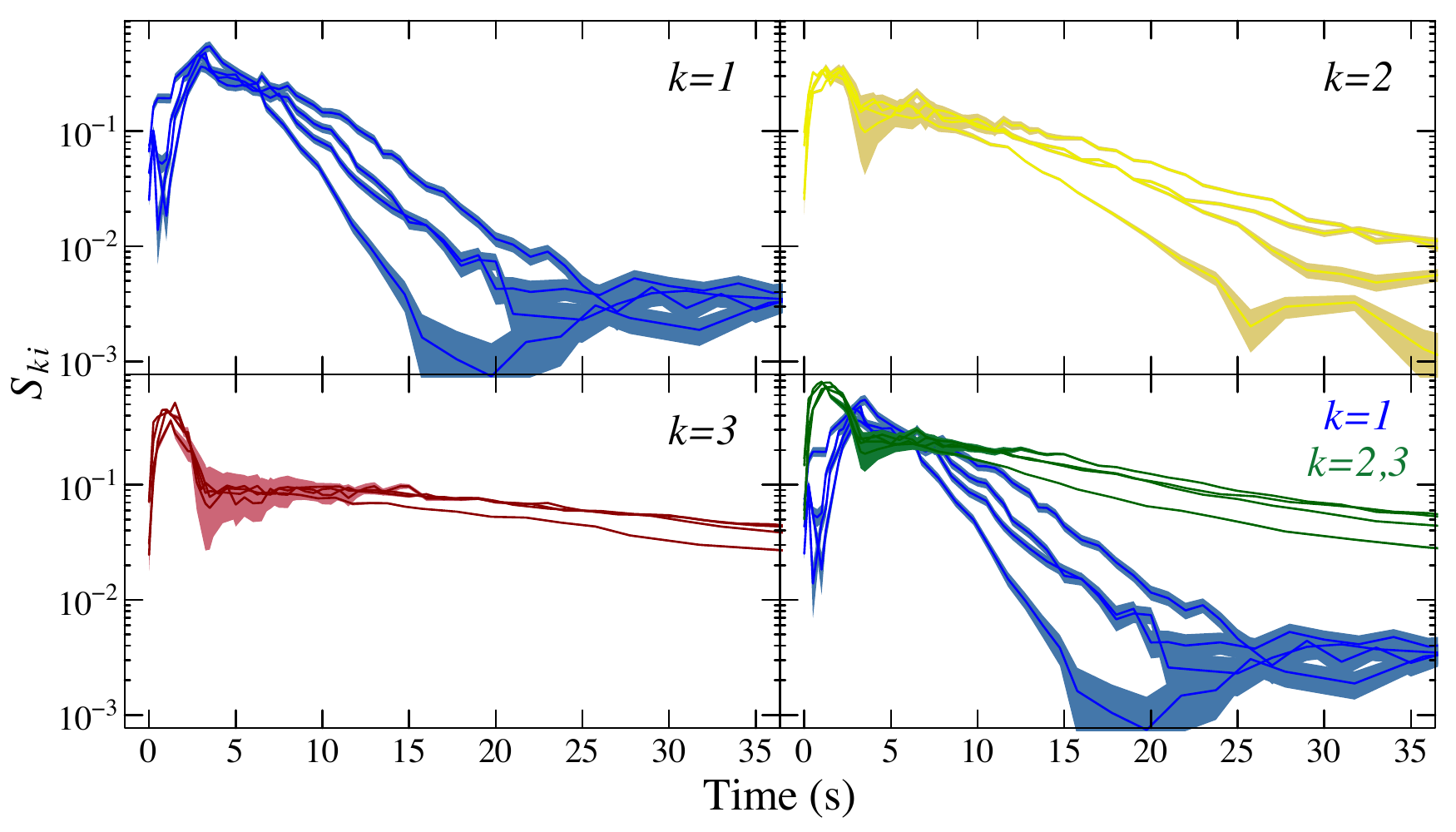}  
\caption{Time evolution of the NMF signals, $S_{ki}$, for the four non-Planckian bursts. The $k=1$ component (which we interpret as the spreading layer) is the dominant component between 3--7~s from the burst onset. In the PRE-phases 0--3~s and the burst tails $>$7~s the sum of the $k=2,3$ components (green lines; which we interpret as the variable temperature blackbody component) dominates. The leveling off of the $k=1$ signal at $\sim$15--25~s corresponds to the time where the spreading layer emission reaches the level observed prior to the burst onset (see text and Fig. \ref{fig:softstatebursts}). }
	\label{fig:nmfevolution}
\end{figure*}

In Fig. \ref{fig:nmfcomponents} we show the spectra ($W_{jk}$) of the three NMF components from multiple runs. The solutions are not unique, which results in a spread of the spectral shape of the individual NMF components. Nevertheless, the resulting NMF components are distinguishable from each other. 
As in the NMF analysis of the 4U 1636--536 superburst in \citet{KKK16}, we interpret the NMF components to arise from a spreading layer varying in normalization ($k=1$) and a variable burst emission varying in normalization and temperature ($k=2,3$).
Note that due to the non-linear effect of the varying blackbody temperature the burst emission gets approximated with two NMF components.
We find that the averaged NMF component $k=1$ from multiple runs is well described with a \textsc{comptt} model, with the seed photon temperature, electron temperature and optical depth of $T_{\rm seed} \approx 1.4$~keV, $T_{\rm e} \approx 3$~keV and $\tau \approx 20$, respectively. 
$T_{\rm seed}$ and $\tau$ are tightly correlated parameters in the fitting, so one of them needs to be fixed to derive constrained confidence intervals.
For example, if $\tau$ is fixed to 12, then $T_{\rm seed} \approx 1.5$~keV (with  the 
90 per cent confidence region being between 1.2 and 1.9~keV) and $T_{\rm e} \approx 3.3\pm0.1$~keV (at 90 per cent confidence). 
We note, however, that the fits have $\chi^2 = 2.7$ for 18 d.o.f., that results from the $k=1$ component having very low flux at energies below $\sim6$ keV.
This causes the lowest flux bins to have very high uncertainties, and thus the low energy slope is not well defined.
Therefore, a blackbody model with $T_{\rm bb} \approx 3.1$~keV can be fitted to the $k=1$ data with similar fit quality.
These parameters are very similar to the \textsc{comptt} model parameters of the variable spreading layer component observed during the persistent emission of NS-LMXBs (including 4U 1608--52, see \citealt{RG06}), who obtained $T_{\rm seed} \approx 1.5$~keV, $T_{\rm e} \approx 3.3$~keV and $\tau \approx 5$.

In Fig. \ref{fig:nmfevolution} we show the time evolution of the NMF signals ($S_{ki}$).
The first component, $k=1$, corresponding to the spreading layer, shows the highest signal strength during 3--7~s from the burst onset, and then decays much faster than the other two components.
On the other hand, both $k=2$ and $k=3$ components show very similar time evolution, which is indicative that they are generated by one spectral component changing both in flux and shape.
Their sum, shown with green lines in the bottom right panel of Fig. \ref{fig:nmfevolution}, is the dominant component in the burst tail, after $\sim$7~s from the burst onsets.

\subsection{Modeling of the persistent emission}

We started the spectral analysis by modeling the persistent emission prior to the burst onsets.
The model we employed consists of four components, \textsc{tbabs} $\times$ (\textsc{diskbb} $+$ \textsc{comptt} $+$ \textsc{gauss}), which were fitted in the 3--20~keV band.
The best fitting parameters for the persistent emission for this model are shown for the eight soft state PRE bursts in Table \ref{tab:persistent}, with the four highly non-Planckian bursts highlighted in bold face. 
The seed photon temperature was fixed to $T_{\rm seed} = 1.5$~keV, the electron temperature to $3.3$~keV and the iron line energy and width to $6.4$ and $0.5$~keV, respectively.
The X-ray fluxes were computed in the $0.01-100$ keV band using the \textsc{cflux} convolution model.

\begin{table*}
\begin{minipage}{170mm}
\caption{
Best-fitting parameters from modeling the persistent spectra prior to the onset of the soft state bursts using \textsc{tbabs} $\times$ (\textsc{diskbb} $+$ \textsc{gauss} $+$ \textsc{comptt}) model.
The first column is the observation ID, where the observations with highly non-Planckian X-ray burst spectra shown in Fig. \ref{fig:softstatebursts} are highlighted with bold face fonts.
Note that the electron temperature and several model parameters were fixed during the fitting (see text).
The Gaussian line equivalent widths were approximately 50~eV. 
}
\centering
\label{tab:persistent}
\begin{tabular}{@{}lcccccccc}
\hline\hline
OBS ID 	& $\Tdbb$ & $\Kdbb$ & $\Fdbb$ & $K_{\rm Fe}$ & $\tau$ & $K_{\rm ctt}$ & $F_{\rm ctt}$ & $\chi^2 / {\rm d.o.f.}$  \\
 	& (keV) &  & ($10^{-9}$ cgs) &  $\times10^{-3}$ &  &  & ($10^{-9}$ cgs) &   \\
\hline
\textbf{30062-01-01-00} 	& $0.881_{-0.013}^{+0.013}$   	& $280_{-20}^{+20}$ 	 	& $3.67_{-0.02}^{+0.03}$ &  $2.1_{-0.5}^{+0.5}$ 	& $3.32_{-0.06}^{+0.06}$  	& $0.115_{-0.002}^{+0.002}$	& $2.213_{-0.013}^{+0.02}$	& $61.6/42$ \\

70058-01-39-00 	& $0.95_{-0.02}^{+0.02}$   		& $170_{-20}^{+20}$ 	 	& $2.99_{-0.06}^{+0.06}$ & $3.3_{-0.6}^{+0.6}$  	& $4.07_{-0.06}^{+0.06}$  	& $0.179_{-0.002}^{+0.002}$	& $3.69_{-0.02}^{+0.02}$	& $42.6/35$\\

70059-01-20-00 	& $0.99_{-0.03}^{+0.04}$   		& $300_{-50}^{+50}$ 	 	& $6.22_{-0.12}^{+0.2}$ & $<1.8$  	& $3.75_{-0.13}^{+0.2}$  	& $0.174_{-0.007}^{+0.006}$	& $3.47_{-0.05}^{+0.03}$	& $38.0/35$\\

\textbf{70059-01-21-00} 	& $1.09_{-0.02}^{+0.02}$   		& $230_{-20}^{+20}$ 	 	& $7.01_{-0.05}^{+0.06}$ & $3.4_{-0.8}^{+0.8}$  	& $3.73_{-0.06}^{+0.06}$  	& $0.229_{-0.004}^{+0.004}$	& $4.57_{-0.02}^{+0.03}$	& $42.9/35$\\

\textbf{70059-03-01-000} & $0.93_{-0.02}^{+0.02}$   		& $330_{-30}^{+30}$ 	 	& $5.21_{-0.06}^{+0.10}$ & $2.3_{-0.7}^{+0.7}$  	& $3.56_{-0.07}^{+0.07}$  	& $0.155_{-0.003}^{+0.003}$	& $3.04_{-0.02}^{+0.03}$	& $57.9/35$\\

93408-01-23-02 	& $0.92_{-0.03}^{+0.04}$   		& $270_{-40}^{+50}$ 	 	& $4.05_{-0.13}^{+0.3}$ &  $3.5_{-1.0}^{+1.0}$ 	& $3.52_{-0.13}^{+0.14}$  	& $0.119_{-0.004}^{+0.004}$	& $2.34_{-0.06}^{+0.03}$	& $40.6/36$\\

95334-01-03-08 	& $0.94_{-0.02}^{+0.02}$   		& $490_{-60}^{+60}$ 	 	& $8.3_{-0.2}^{+0.2}$ &  $4.7_{-1.5}^{+1.5}$ 	& $3.39_{-0.09}^{+0.09}$  	& $0.243_{-0.006}^{+0.006}$	& $4.70_{-0.05}^{+0.11}$	& $32.7/36$\\

\textbf{96423-01-11-01} 	& $0.90_{-0.02}^{+0.02}$   		& $410_{-30}^{+40}$ 	 	& $5.84_{-0.11}^{+0.11}$ & $1.8_{-0.7}^{+0.7}$ 	& $3.41_{-0.06}^{+0.06}$  	& $0.171_{-0.003}^{+0.002}$	& $3.31_{-0.02}^{+0.02}$	& $30.9/36$\\
\hline

\end{tabular}
\end{minipage}
\end{table*}

The model fits the data rather well, with the best fitting \textsc{comptt} optical depths consistent with each other among the persistent spectra with a mean of $\tau \approx 3.59$, although for observations of IDs 30062-01-01-00 and 70059-03-01-000 the model is not statistically acceptable.
The \textsc{diskbb} normalizations, on the other hand, allows us to estimate the inner disc radius. 
We obtain a mean value of $R_{\rm in} \approx 10.7$~km (minimum and maximum values being $R_{\rm in,min} \approx 8.0$ and $R_{\rm in,max} \approx 13.5$~km) using equation (1) of \citet{GD02}, and assuming the distance of $d=3.6$~kpc, the colour-correction factor of $f_{\rm c}=1.8$,  inclination of $60\deg$ and the correction term for the inner disc boundary condition of $\eta = 2.7$.
Thus, the disc component is consistent of being extended to the NS surface.
We also see that the \textsc{comptt} component carries approximately 35--55 per cent of the estimated bolometric flux.
Importantly, the \textsc{comptt} parameters are in agreement with the NMF decomposition results of the bursts as well as with the frequency-resolved spectroscopic measurements of the persistent emission made by \citet{RG06}.

\subsection{Burst modeling}

In addition to the standard blackbody fits where the persistent emission is removed as a background, we also modeled all the burst spectra with a two-component model using the aforementioned two methods (see Section 2).
In the first method we removed the background as before, but described the X-ray burst emission with the blackbody model and used the Comptonization model \textsc{comptt} as an additional component.
In the second method we did not remove the persistent emission as a background, but instead we fixed the \textsc{diskbb} and \textsc{gauss} components to the values obtained prior to the bursts, and allowed only the \textsc{comptt} component to vary.

After series of tests, we decided to fix the \textsc{comptt} model parameters to the average best fitting values of the variable spreading layer component observed during the persistent emission given Table \ref{tab:persistent}, and only allow the flux (i.e. model normalization) to vary.
We also tested various other $\tau$ and $\Te$ values, even allowing both of them to vary, and found that the results do not change qualitatively for the particular choice of these parameters.
The results did not differ either between the different fitting methods in a qualitative sense.
That is, the best fitting parameters and their evolution during the bursts were largely unaffected even if we modeled, or subtracted, the persistent emission as a background. 
The only difference in the second method was that during the bursts the \textsc{comptt} component could freely result in fluxes that were smaller than prior to the onsets of the bursts shown in Table \ref{tab:persistent}.
We decided to display the parameter evolution using the second method, but we re-iterate that the other modeling methods and slightly different choices in fixing parameters yield comparable results.

Finally, we modeled the burst spectra with only the \textsc{comptt} model, to study the possibility that occasionally the entire NS is covered beneath the spreading layer. For these fits we show only the goodness of fit measure in the subsequent figures. 

\begin{figure}
\centering
\includegraphics[width=\linewidth]{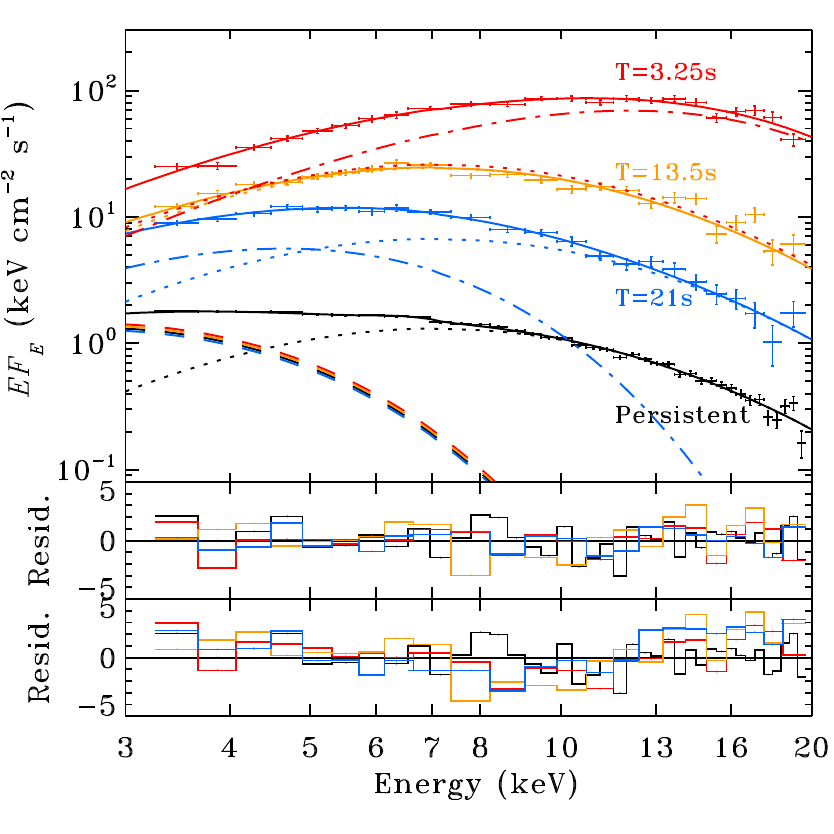}
\caption{Example energy spectra of the burst and persistent emission from OBSID 70059-03-01-000.
The solid line shows the total model spectrum, the dotted line shows the variable \textsc{comptt} component, the dashed line shows the non-variable \textsc{diskbb} component, and the dot-dashed line shows the burst blackbody emission.
Note that for the burst spectrum at 13.5~s from the burst onset the spectrum can be fitted only with the \textsc{comptt} component.
The middle panel shows the residuals (in units of $\sigma$) from the variable \textsc{comptt} $+$ (\textsc{bb} and/or \textsc{diskbb}) fits.
The bottom panel shows the residuals for the \textsc{bb} fits, which systematically underpredicts the burst emission above $\approx13$~keV. 
}
\label{fig:spectra}
\end{figure}

\subsection{Burst parameter evolution}

Three example burst spectra are shown in Fig. \ref{fig:spectra} for the burst occurring during OBSID 70059-03-01-000, together with the persistent spectrum shown with black lines.
The dotted lines mark the spectrum of the \textsc{comptt} component, except in the spectrum at 13.5~s from the burst onset, in which \textsc{comptt} is the only spectral component and therefore marked as a solid line.
During this time, a blackbody spectrum has almost an identical shape when compared to the \textsc{comptt} model shape (the \textsc{comptt} component in the spectrum at 3.25 s overlaps almost exactly the data points of the spectrum at 13.5 s), making it difficult to constrain the parameters of the blackbody model. 
The addition of the \textsc{comptt} component removes the wavy residuals of the blackbody fits.

\begin{figure*}
	\centering
	\begin{tabular}{c c}
		\includegraphics[width=0.43\linewidth]{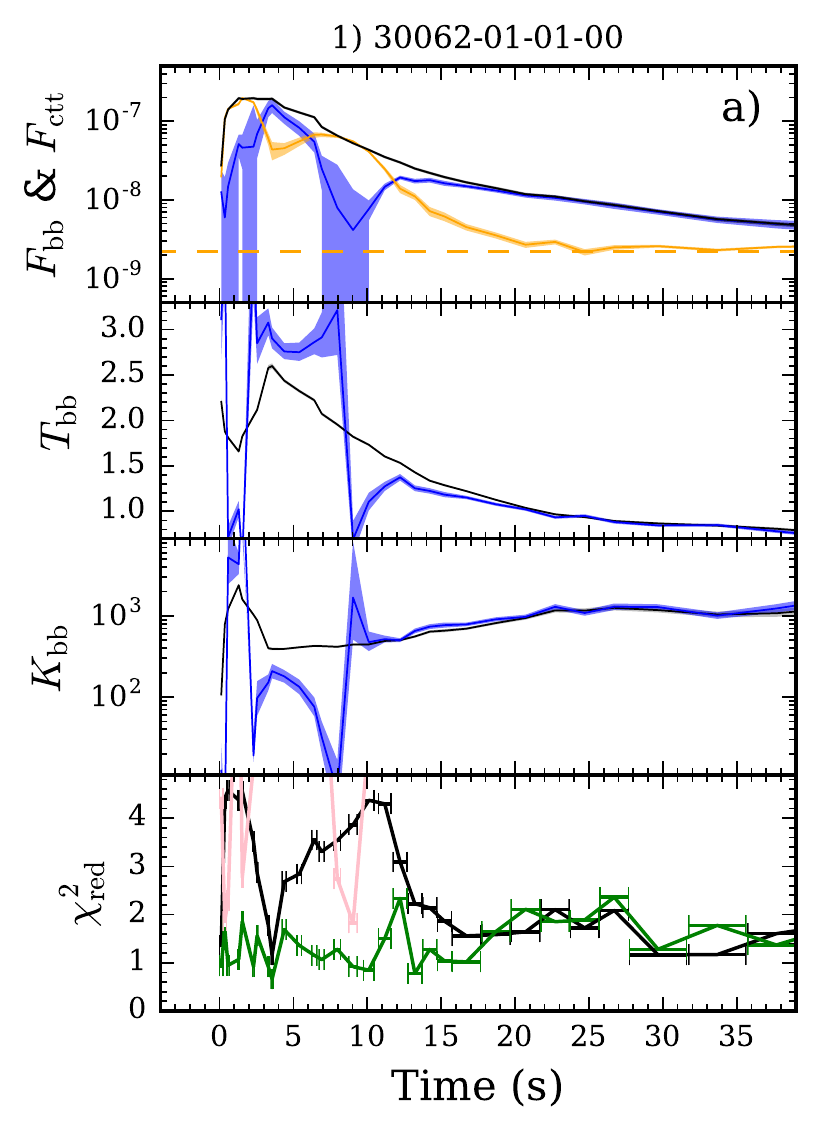}  & \includegraphics[width=0.43\linewidth]{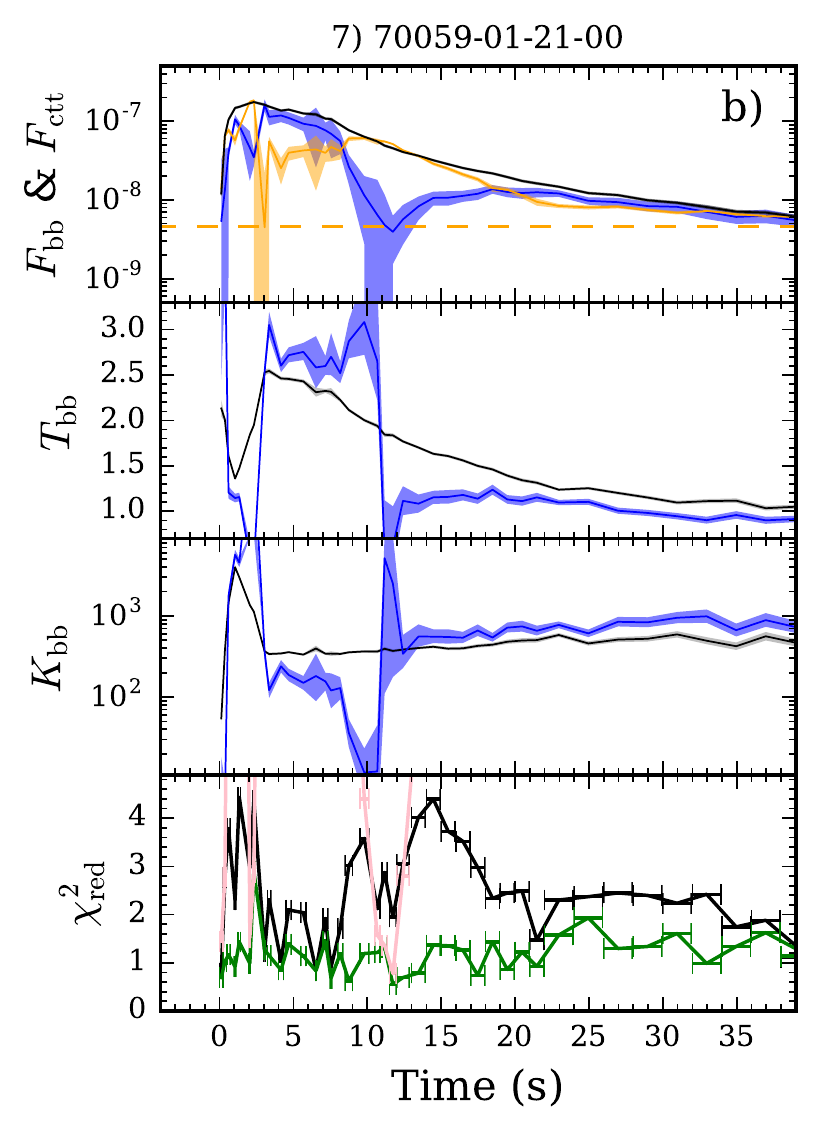}  \\

		\includegraphics[width=0.43\linewidth]{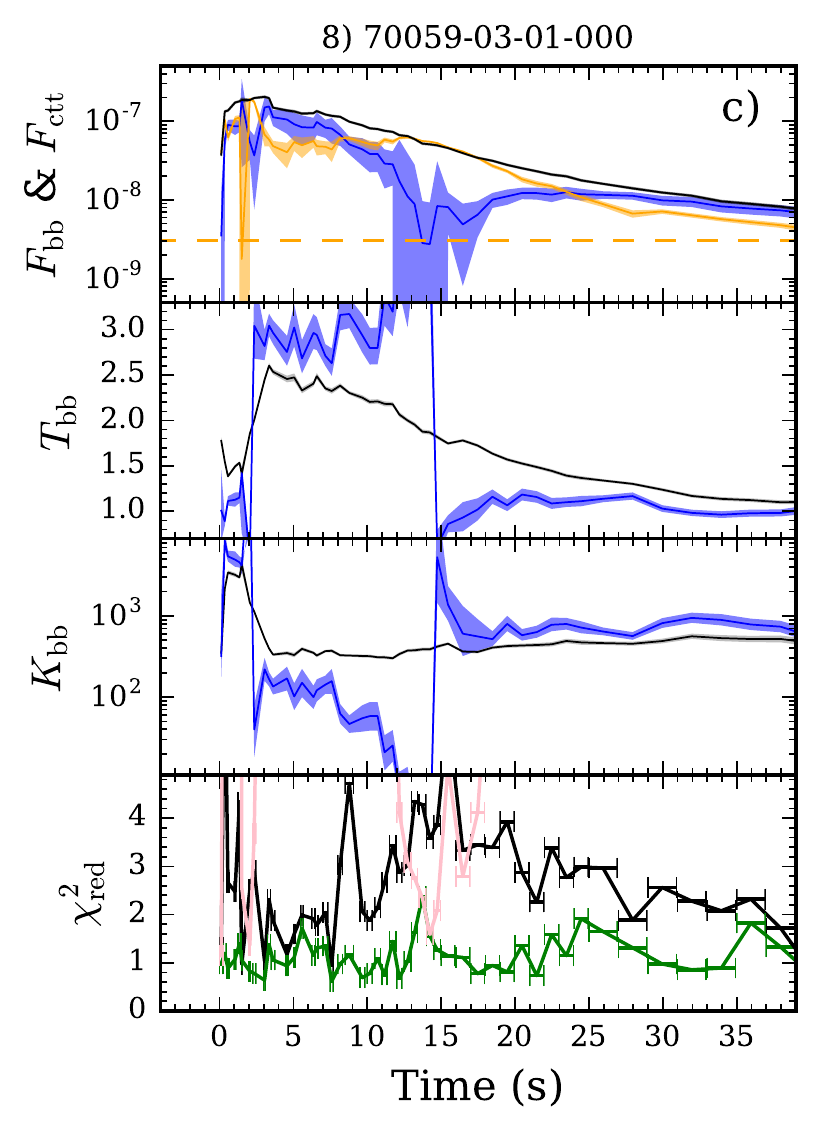}  & \includegraphics[width=0.43\linewidth]{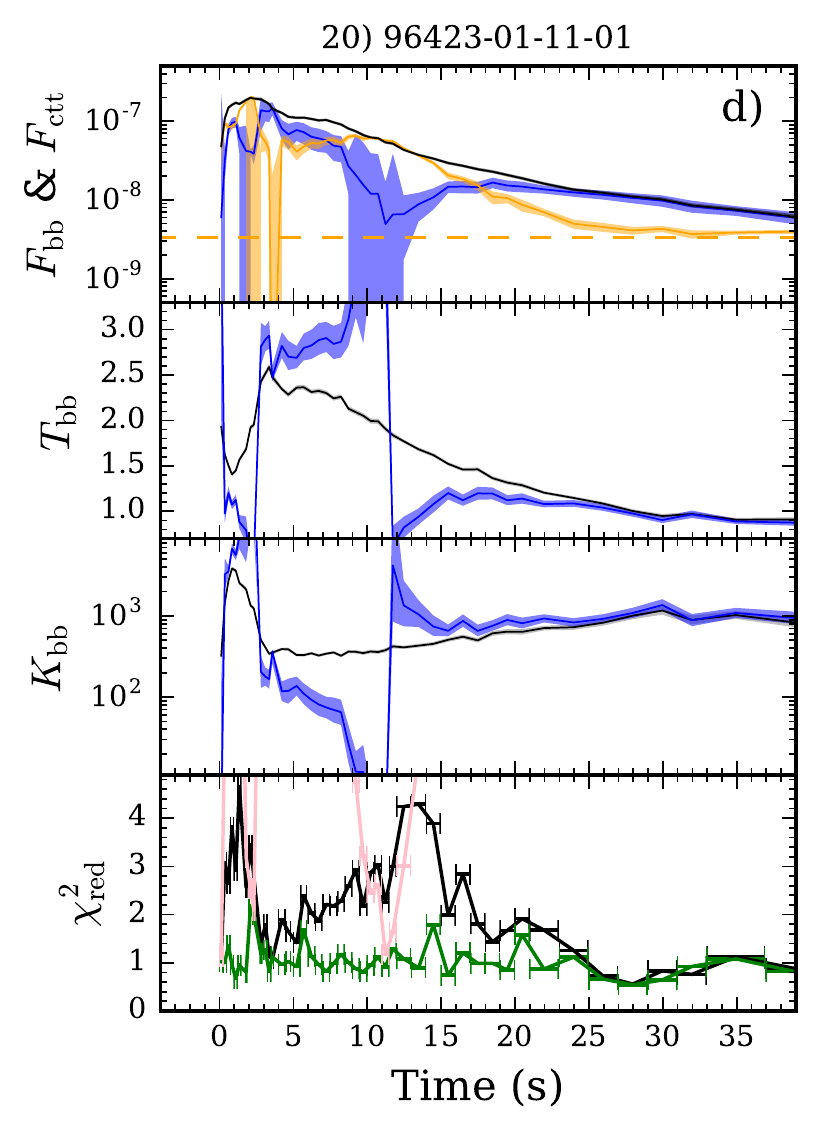} 	
	\end{tabular}
	\caption{Time evolution of the model parameters fitted to the 4U 1608--52 PRE burst spectra during the soft spectral state. Black lines show the time evolution of the blackbody model parameters from a classic blackbody fit. Blue and orange lines show the evolution of the model parameters of the two-component model consisting of a blackbody (\textsc{bb}) and Comptonization (\textsc{comptt}) components, respectively. \textsc{comptt} parameters were fixed to the averaged value fitted to the spectra observed prior to the bursts, and only \textsc{comptt} normalization (i.e flux) was left free to vary. The \textsc{comptt} flux level prior to the burst is shown with a dashed line. In the bottom panels, the green, pink and black lines show the reduced chi-squared values for the two-component, pure Comptonization (i.e. no burst blackbody component at all) and pure blackbody fits, respectively. }
	\label{fig:softstatebursts}
\end{figure*}

The time evolution of the spectral parameters for the \textsc{xspec} spectral fits are shown in Fig. \ref{fig:softstatebursts} for the four highly non-Planckian bursts. 
The black lines show the results from the blackbody fits (grey bands indicate 1$\sigma$ errors for each parameter).
The blue and yellow lines (and 1$\sigma$ error bands colored with lighter shades) show the results of the blackbody and \textsc{comptt} components, respectively, with the reduced $\chi^2$ of the fits displayed with a green line in the bottom panel.
The bottom panel also shows the $\chi^2_{\rm red}$ for the case where the entire burst spectrum is modelled only with the \textsc{comptt} model (pink lines).

\begin{figure*}
\centering
\includegraphics[]{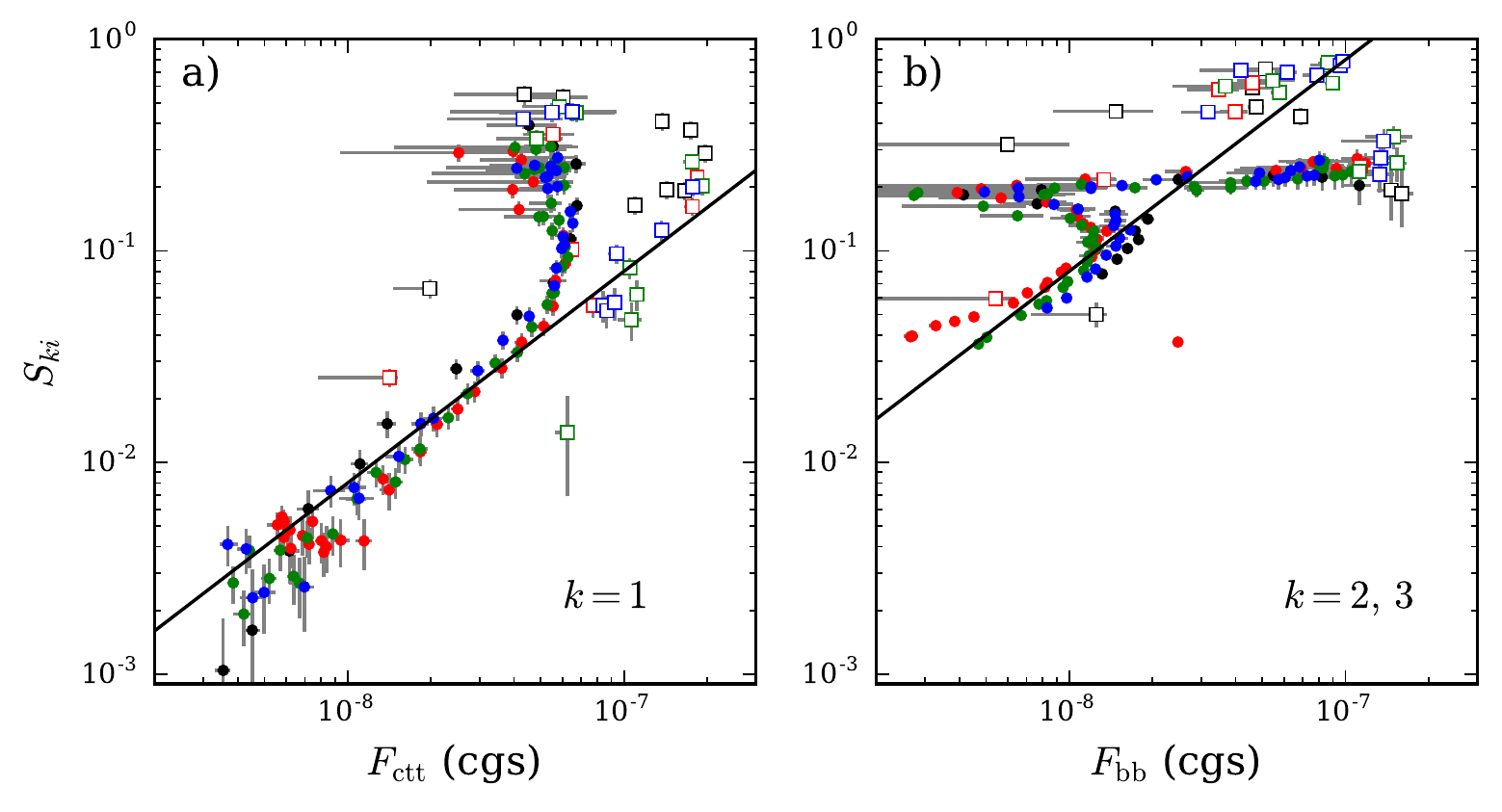}
\caption{Comparison between the NMF signals to the fluxes obtained from the spectral fits.
(a) The signal for the $k=1$ component, $S_{ki}$, versus the \textsc{comptt} flux, $F_{\rm ctt}$. 
(b) The signal for the $k=2,3$ component versus the \textsc{bb} flux, $F_{\rm bb}$.
The black, red, green and blue colours correspond to the four different bursts shown in Fig.~\ref{fig:softstatebursts}(a), (b), (c) and (d), respectively. 
Squares and circles highlight spectra that occur in the PRE-phase and the cooling tail, respectively.
A one-to-one relationship holds for fluxes below $F_{\rm ctt} \lesssim 6\times10^{-8}\,\ergcms$, indicating that the direct spectral modeling works well in this range. In the end of the PRE-phase, and the early cooling tail the spectral fits and the NMF decomposition yield highly contradictory results.
}
\label{fig:NMF_vs_comptt}
\end{figure*}

The \textsc{comptt + bb} model provides a highly significant improvement to the fit  for these 4 bursts with respect to the \textsc{bb} model fit.
The fits indicate that the \textsc{comptt} component is the dominant spectral component particularly in the beginning of the cooling phase of the burst, often being consistent of carrying 100 per cent of the burst flux.
It is during these times that the parameters of the blackbody component become unconstrained in the \textsc{xspec} fits.
This result is different with respect to the NMF decomposition, where we saw that the $k=1$ component (corresponding to the \textsc{comptt} model) signal $S_{1i}$ is comparable to the summed $k=2,3$ one. 
In fact, by comparing the parameter evolution to the NMF decomposition in Fig. \ref{fig:nmfevolution} we see that the \textsc{xspec} fitting results are very different in the brightest parts of the bursts.
For example the PRE phase seems to be dominated by the \textsc{comptt} component, whereas the NMF decomposition shows that the highest signal comes from the varying blackbody components ($k=2,3$).
Here it is important to keep in mind the major difference between the two methods:
Fitting spectra in \textsc{xspec} is done individually for each timestep whereas the NMF decomposition tries to minimize a global cost function spanning over the whole burst.
This difference is discussed more in the Sect.~4.

While in the blackbody  fits -- and in the NMF decomposition -- the parameter evolution is smooth,
we see that in the two component fits  the burst temperature shows much more rapid swings from cold to hot phases (and back) during the bursts.
As we discuss below, these rapid changes are likely due to fitting degeneracies and do not correspond to real changes of the burst temperature.
During these temperature swings two noticeable things occur.
First, we see that the swings coincide with the times when the \textsc{comptt}-only fits provide a much better fit than the blackbody model (compare pink and black $\chi^2_{\rm red}$ curves).
Secondly, in some cases the swings occur a couple of seconds after the blackbody temperature has a clear kink in its evolution, typically about 7 seconds after the burst onsets.\footnote{In some decompositions with different \textsc{comptt} parameters these two times coincide.} 
That is, before the kink occurs the blackbody temperature falls down much more slowly than after it.
It is only during this ``slow cooling'' stage that we can observe the constant blackbody radii in the blackbody fits, which are not predicted by the atmosphere models \citep{SPW12}. 

To highlight the differences between the two decompositions, we compare the strength of the $k=1$ NMF signals, $S_{ki}$, and the \textsc{comptt} component fluxes, $F_{\rm ctt}$, in Fig.~\ref{fig:NMF_vs_comptt}(a).
Similarly, in Fig.~\ref{fig:NMF_vs_comptt}(b), we compare the $k=2,3$ NMF signals to the blackbody component fluxes, $F_{\rm bb}$.
There is a clear one-to-one relationship for fluxes below $F_{\rm ctt} \lesssim 6\times10^{-8}\,\ergcms$ for all four bursts (highlighted with different colours). Also, during the PRE-phase of these bursts that lasts between $3.25$--$3.5$~s (shown by squares), the spectra show roughly similar $F_{\rm ctt}$ as compared to the $F_{\rm ctt}$--$S_{ki}$ trend line later in the cooling tail.
Similar behavior is also seen in Fig.~\ref{fig:NMF_vs_comptt}(b).
However, the differences are highly notable in the end of the PRE-phase and in the early cooling stage where according to the \textsc{comptt + bb} fits the \textsc{comptt} component dominates. 
The model fits simply do not agree with the NMF decomposition results in this range, which probably arises from the spectral degeneracy of the {\sc comptt} and blackbody components.  

\begin{figure*}
\centering
\begin{tabular}{c c}
\includegraphics[width=0.43\linewidth]{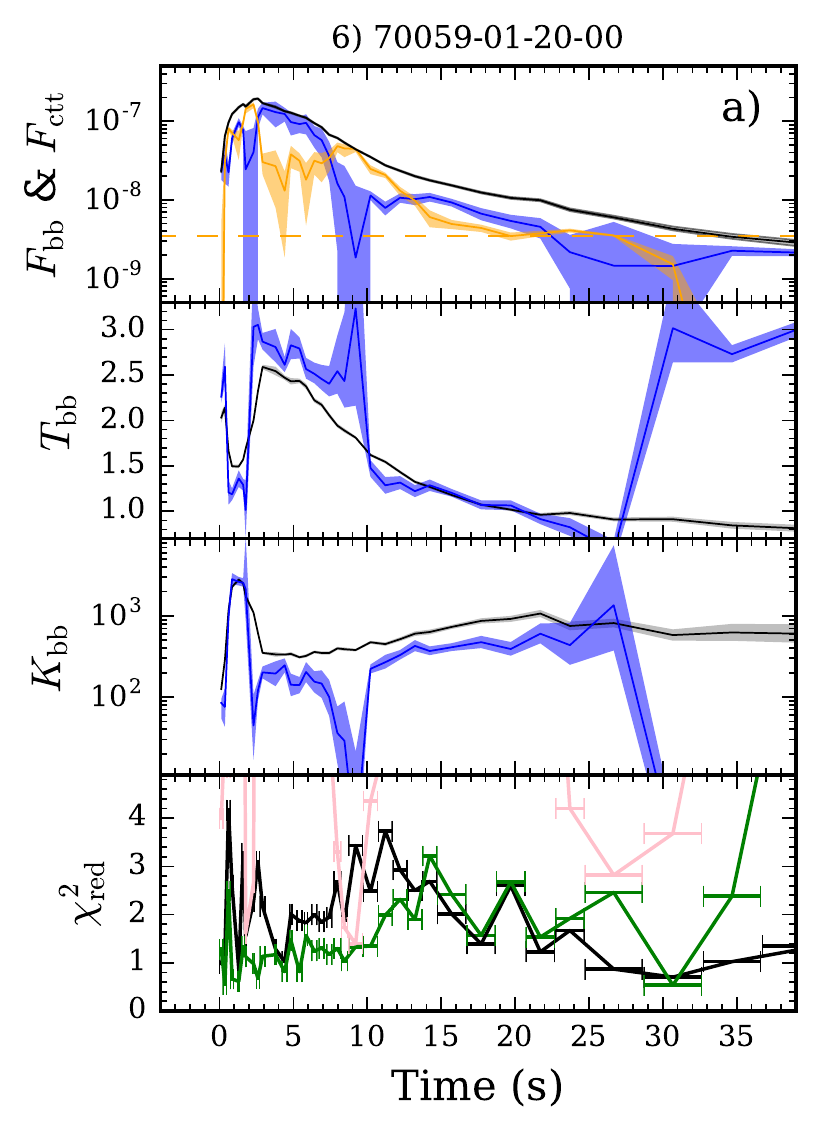} & \includegraphics[width=0.43\linewidth]{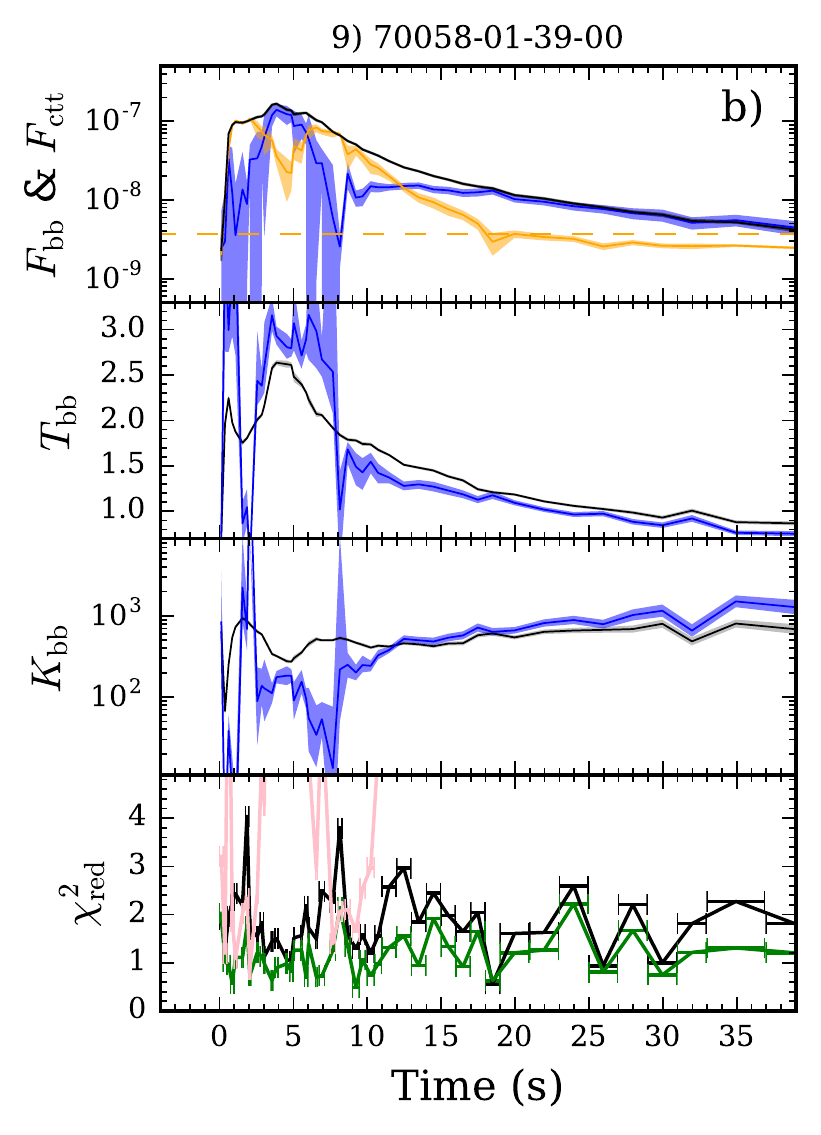} \\

\includegraphics[width=0.43\linewidth]{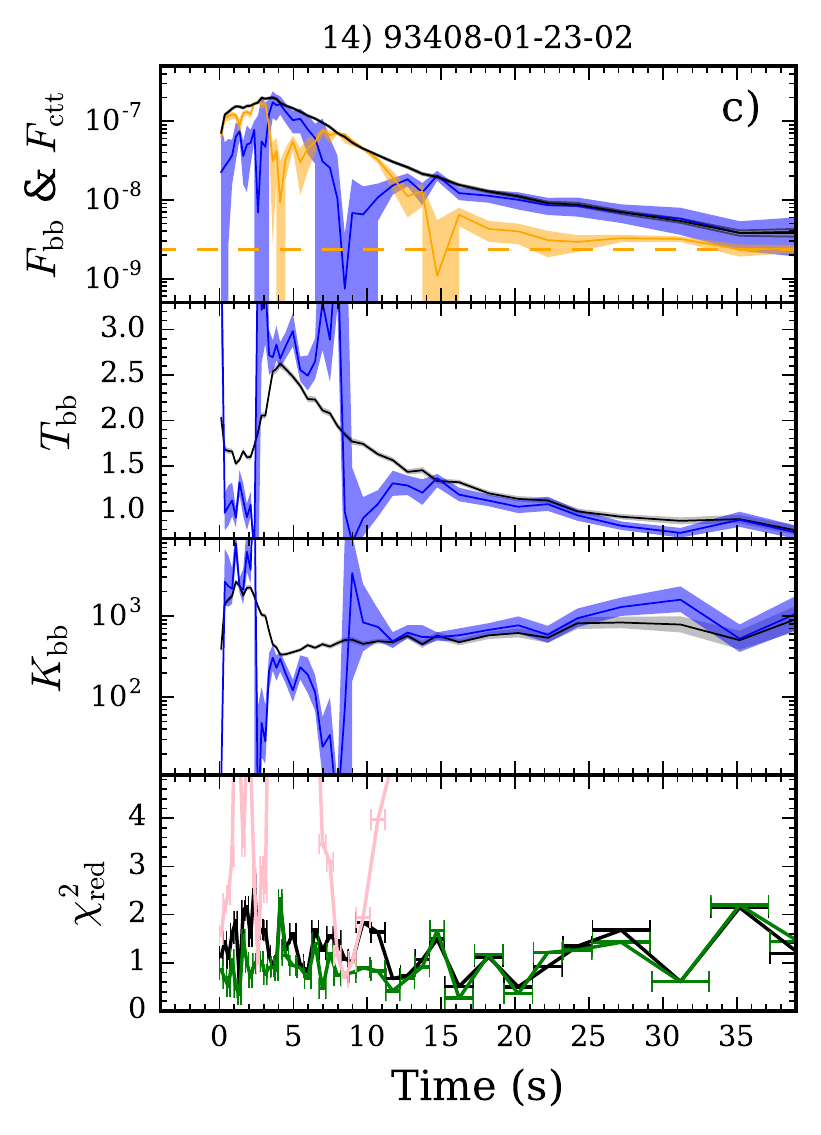} & \includegraphics[width=0.43\linewidth]{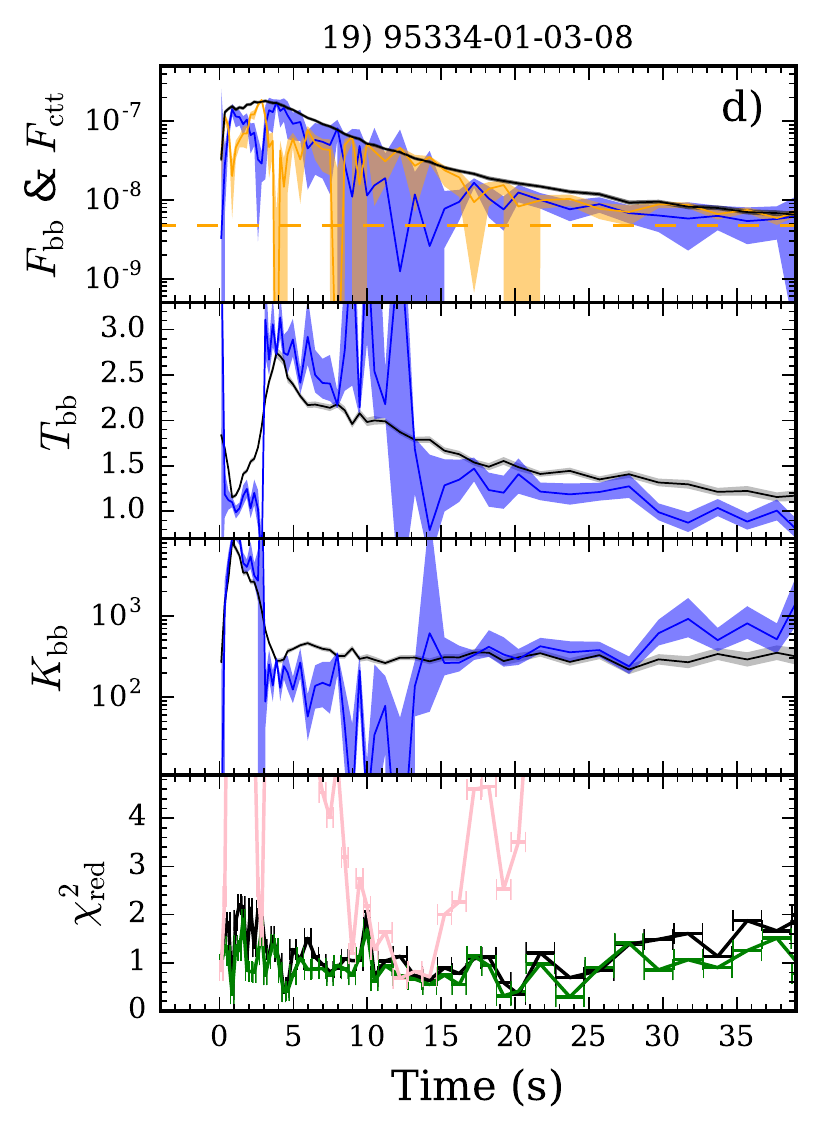}   \\

\end{tabular}
\caption{Same as  Fig.~\ref{fig:softstatebursts}, but for four other  soft-state PRE bursts. }
\label{fig:softstatebursts2}
\end{figure*}

The results from the other four PRE-bursts in the soft state show similar behavior, albeit the presence of the second spectral component is less clear as can be seen in Fig.~\ref{fig:softstatebursts2}.
The model parameter evolution in the panels Fig.~\ref{fig:softstatebursts2}(a)--(d) show that the burst spectra is fitted adequately (except the first 10--15 s) with a pure blackbody model and, consequently, the \textsc{comptt} component becomes less constrained and unnecessary.
It is worth highlighting, however, that very similar trends can be observed as compared to the bursts in Fig.~\ref{fig:softstatebursts}, thus the same physical processes are likely to take place.

\begin{figure*}
\centering
\begin{tabular}{c c}
\includegraphics[width=0.43\linewidth]{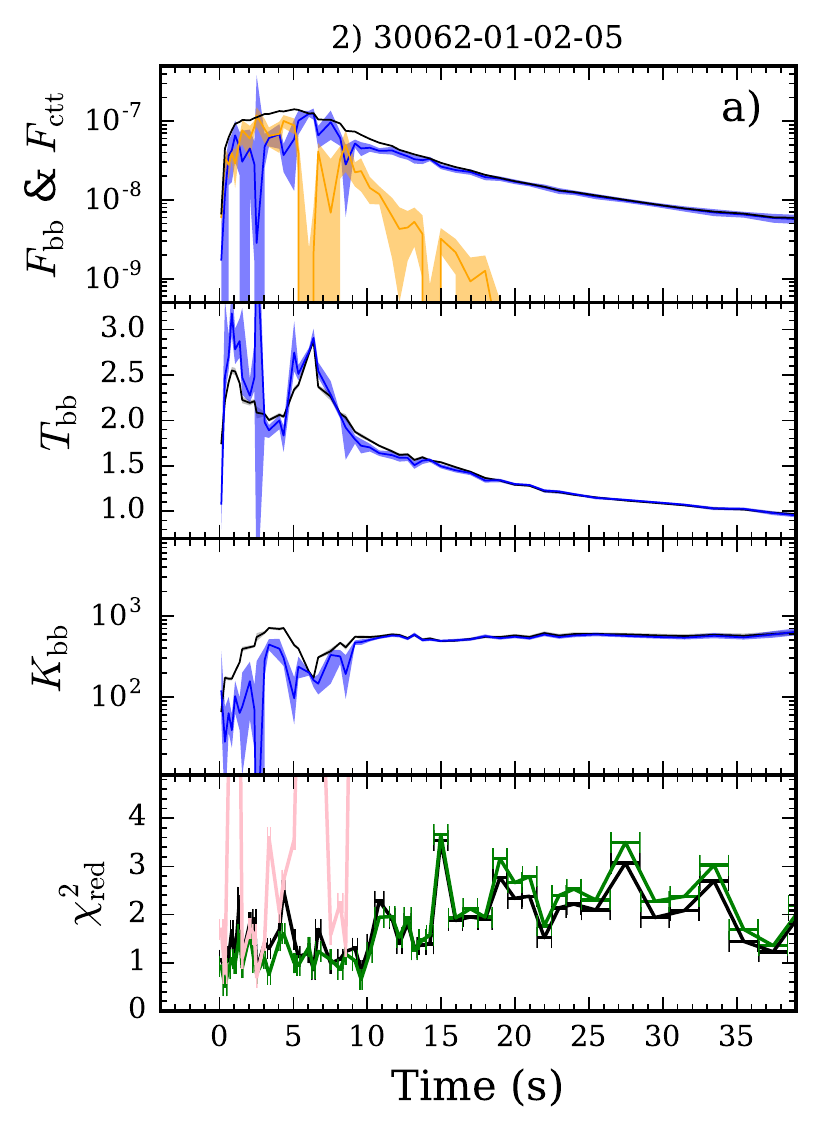}  & \includegraphics[width=0.43\linewidth]{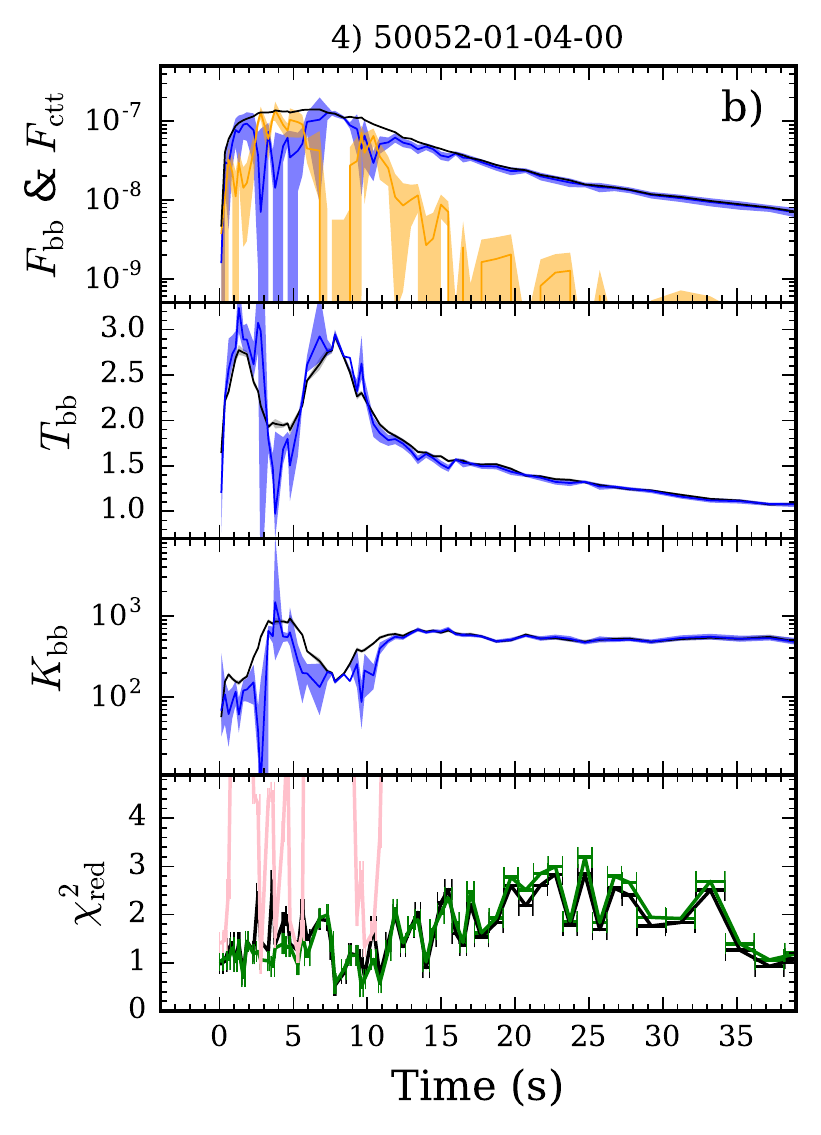}  \\

\includegraphics[width=0.43\linewidth]{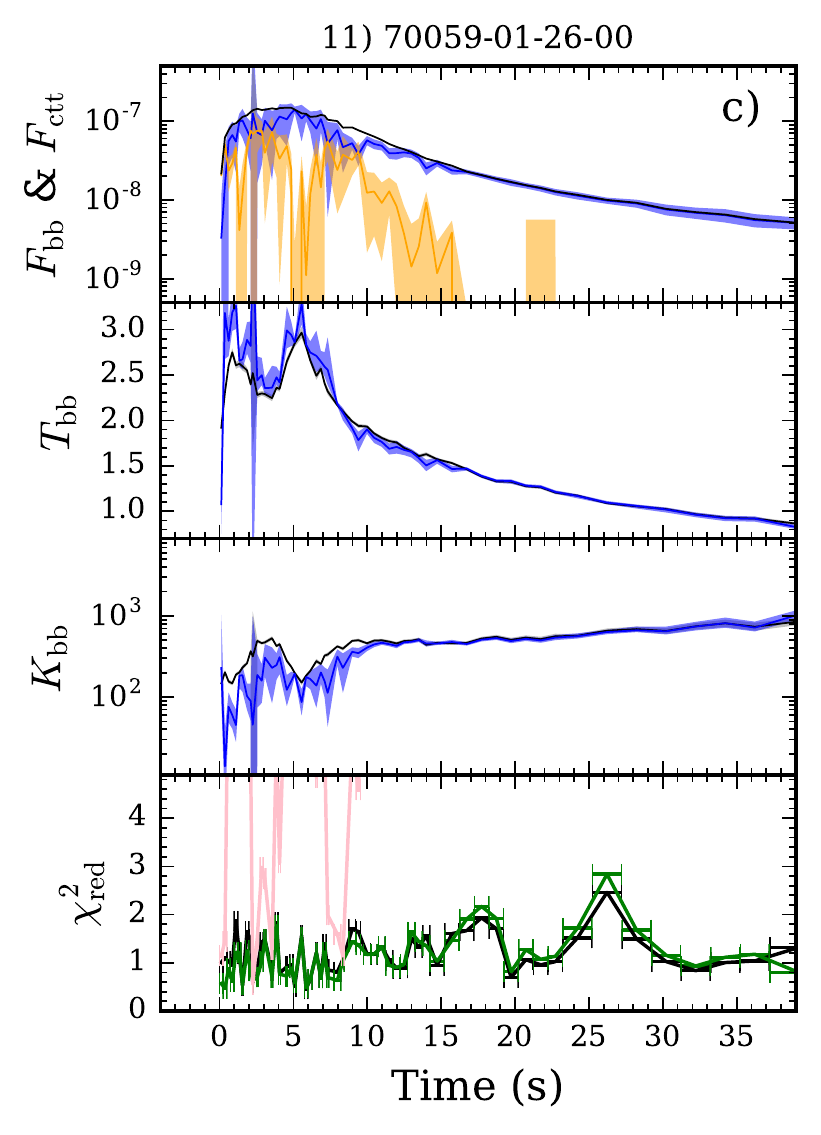}  & \includegraphics[width=0.43\linewidth]{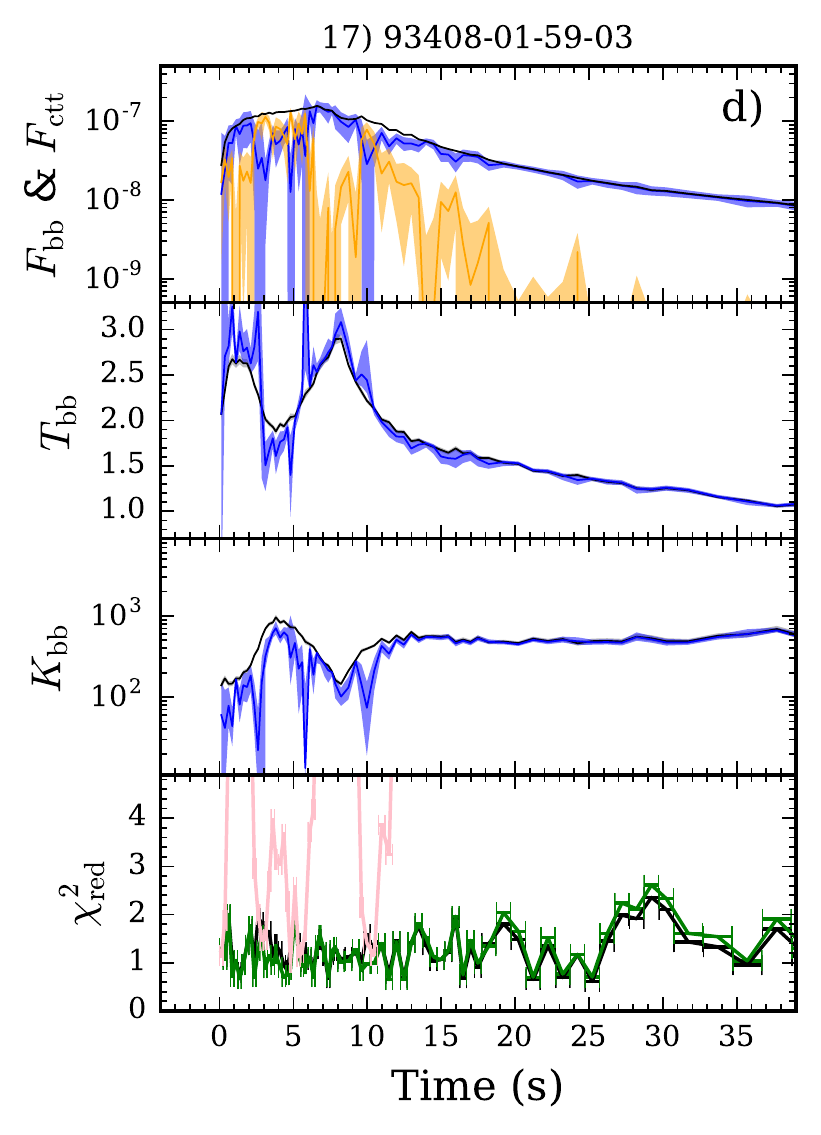} 

\end{tabular}
\caption{Same as  Fig.~\ref{fig:softstatebursts}, but for four hard-state PRE-bursts that were used by \citet{PNK14} to determine the NS mass and radius of 4U 1608--52. 
Note that the pre-burst spectrum has been subtracted as a background. The burst spectra are well fitted with a blackbody model, and adding the \textsc{comptt} component does not improve the fits, not even in some burst tails where the blackbody fits resulted in $\chi^2_{\rm red} \approx 2$.
}
\label{fig:hardstatebursts}
\end{figure*}

\begin{figure*}
\centering
\begin{tabular}{c c}
\includegraphics[width=0.49\linewidth]{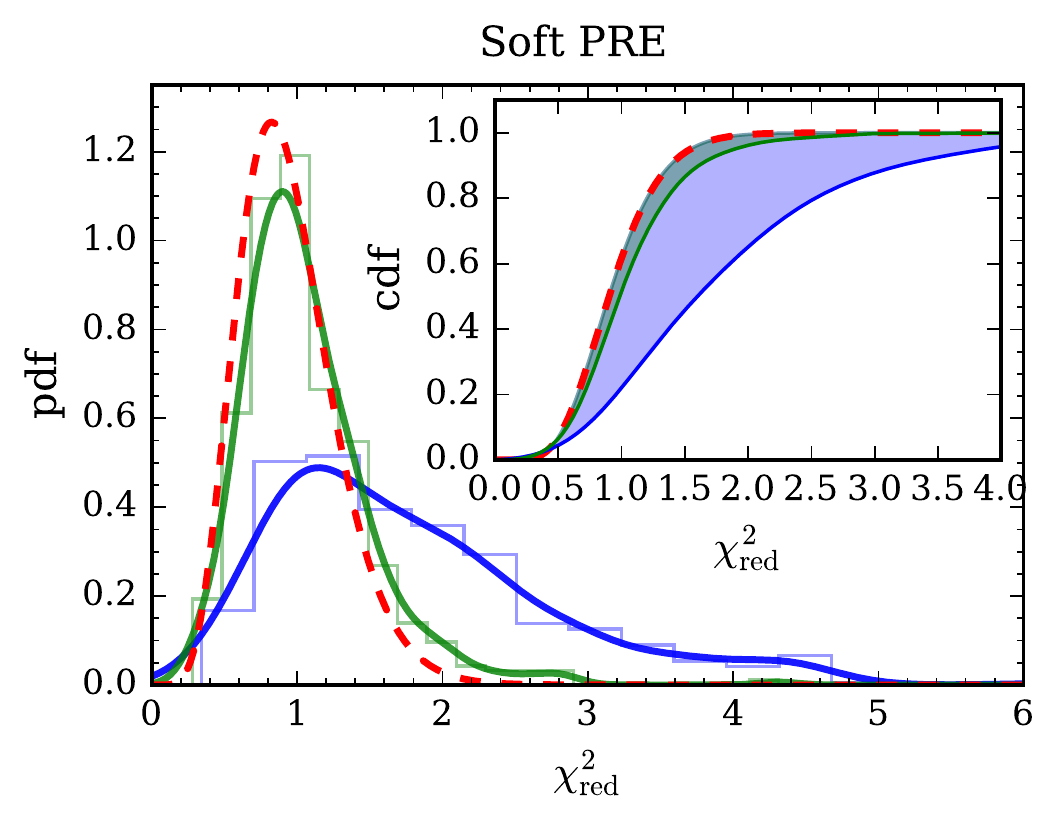}  & \includegraphics[width=0.49\linewidth]{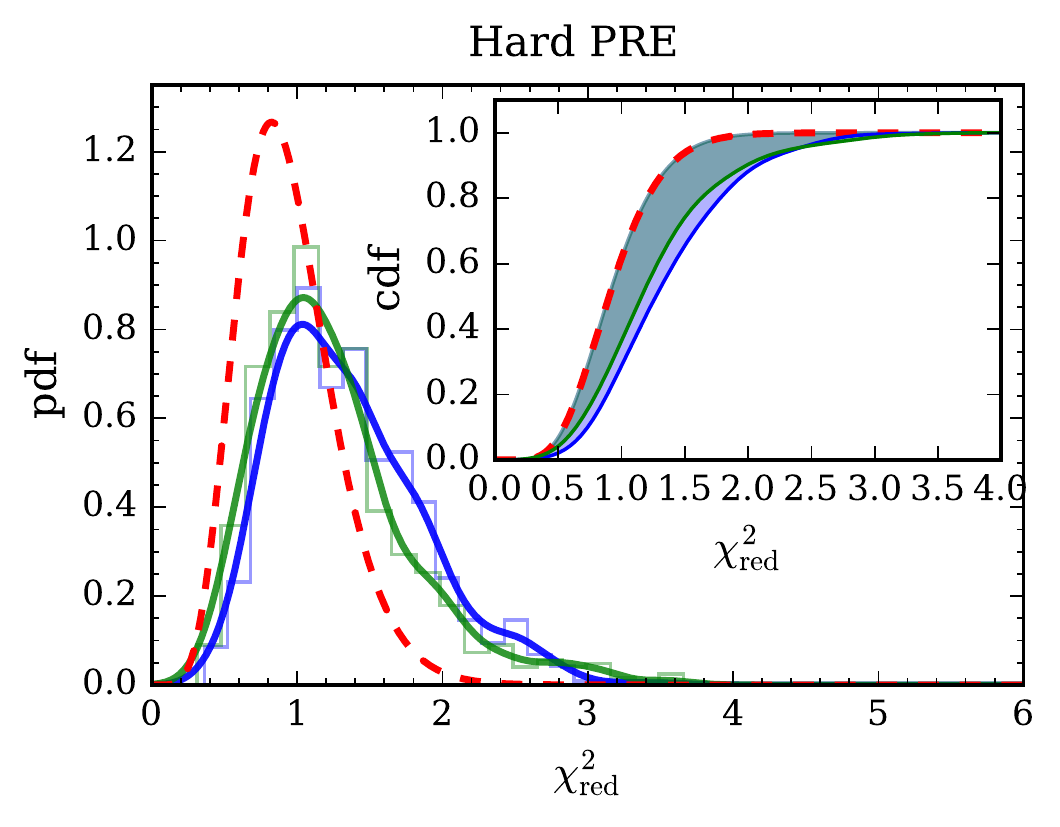} \\
\end{tabular}
\caption{Reduced $\chi^2$ distributions for the soft-state (left panel) and the hard-state (right panel) PRE burst spectra. 
The step-line shows the underlying histogram of the distribution while the thick solid lines correspond to Gaussian kernel density estimates of the distribution.
Green and blue lines show the estimates for the \textsc{bbodyrad}$+$\textsc{comptt} and \textsc{bbodyrad} models, respectively.
Red dashed lines show the expected theoretical distributions (for d.o.f. equal to 17, a typical value for the observed spectral bins).
Insets in the upper right corners show the cumulative distributions where the deviations from the theoretical distribution are highlighted with green and blue shadings.
All the distributions are normalized so that the area encapsulated by the curve is unity.
}
\label{fig:chired}
\end{figure*}

We also performed similar analysis for the hard state bursts.
In Fig. \ref{fig:hardstatebursts}, we show the time evolution of the model parameters taken from four bursts that were used by \citet{PNK14} to derive the NS mass, radius and distance of 4U 1608--52.
These fits show clearly that in the PRE phase and during the early cooling phases the blackbody model fits the data well.
Only at lower fluxes we see values of $\chi^2_{\rm red} \approx 2$, and more importantly the fits are not improved by adding the \textsc{comptt} model.
This difference is further highlighted in Fig. \ref{fig:chired}, where we show the $\chi^2_{\rm red}$ distributions for the soft (left panel) and hard state bursts (right panel).
We can see that for the soft state bursts the addition of the \textsc{comptt} component brings the observed distribution close to the expected one, whereas in the hard state the bad fits during the end of the cooling tails are not removed.

\section{Discussion}

The non-Planckian spectra seen during the soft state PRE-bursts of 4U 1608--52 are caused by another spectral component that is well described by an optically thick Compotonized emission that we model using \textsc{comptt}.\footnote{Note that non-Planckian spectra from 4U 1608--52 have been observed also with TENMA \citep{NDI89}.}
The results indicate clearly that the parameters of the \textsc{comptt} component during the X-ray bursts share many similarities with the parameters of the spreading layer that forms in the accretion disc-NS boundary, both from theoretical perspective \citep{IS99,IS10,SP06}, with respect to previous observations of the persistent emission \citep{GRM03,RG06,RSP13}, as well as the 4U 1636--536 superburst data \citep{KKK16}.
The spreading layer model essentially suggests that the layer -- that spreads from the NS equator (disc mid-plane) towards the NS rotational poles --  is levitating above the NS surface, and the bulk of the emission is released in two bands located above the equator and up to a latitude higher up that is determined by the mass accretion rate.
Each latitude in these emission bands emit quasi-thermal emission at its local Eddington limit \citep{SP06}.
The model predicts a roughly constant (Eddington) temperature emission from each latitude, and that the latitudinal width is solely determined by the emitted flux, which is proportional to mass accretion rate.
Given that the spreading layer spectral shape is constant irrespective of the emitted flux, the model thus naturally explains why during the soft state (i.e. banana branch) the hard X-ray colour of 4U 1608--52 (and other ``atoll'' sources) remains constant over a large range of observed fluxes \citep{GRM03,RG06,RSP13}.

The spreading layer carries a little less than 40 per cent of the persistent flux in the cases we studied, consistent with the findings of \citet{TSM11} who analyzed the persistent emission of 4U 1608--52 using \textit{RXTE} data.
However, as can be seen in Fig. \ref{fig:spectra}, the spreading layer actually dominates the persistent emission in the $3$--$20$~keV spectral band of the PCA instrument, particularly above $\sim 7$~keV.
Therefore, the method of multiplying the entire persistent spectrum with a constant \citep{WGP13,WGP15} -- which was found to be spectral state dependent in 4U 1608--52 \citep{JZC14} -- is indeed very similar to our approach.
On the other hand, our interpretation of the whole phenomenon is very different.
Similarly to the superburst of 4U 1636--536 \citep{KKK16}, the non-parametric NMF decomposition clearly shows that only the spreading layer component is variable during the bursts and the disc component that dominates at lower energies remains constant.
As the disc likely extends down to the innermost stable circular orbit (or the NS surface), it should get hotter if the prevailing mechanism for the persistent emission change is an increase of the mass accretion rate through the Poynting-Robertson drag mechanism, as suggested by  \citet{WGP13,WGP15}.
This heating and the subsequent cooling of the disc during the burst tail, is not seen in the NMF analysis nor is it needed to describe the spectral changes.

An alternative interpretation was proposed by \citet{KKK16} based on the 4U 1636--536 superburst data.
The observed variations of the \textsc{comptt} component can be interpreted as changes in the latitudinal width of the spreading layer in the course of the X-ray burst.
The layer is proposed to be levitating above the NS surface, because it is being pushed outwards by the radiation pressure approaching the Eddington limit \citep{IS99,IS10,SP06}.
Clearly, during the X-ray burst, the increased emission from the burst must travel through this layer near the NS equator providing additional pressure support.
We propose that -- as a consequence of the increased emission -- the layer is being pushed towards higher latitudes, perhaps completely engulfing the NS as the burst emission approaches the Eddington limit.
Hence, rather than seeing a bursting NS atmosphere near the burst peak, we see a large fraction of the burst emission passing through the spreading layer, and subsequently, as the burst flux decays the spreading layer latitudinal width decreases, ultimately returning to its original state observed prior to the X-ray burst.

Such phenomenon was speculated to exist by \citet{SPR11} -- and also by \citet{PNK14} and \citet{KNL14} -- to describe why the blackbody radii immediately after the photospheric touchdown were constant in soft state bursts.
The blackbody radii of the undisturbed NS atmosphere are not expected to be constant, given that they depend on the colour correction factor as $\Rbb \propto f_{\rm c}^{-2}$, as $f_{\rm c}$ has a strong dependency on the emitted flux, especially near the Eddington limit \citep{SPW11,SPW12}.
The models predict a very characteristic increase of the blackbody radii in the beginning of the cooling stage, almost identical to what is seen in the hard state bursts of 4U 1608--52 (see Fig. \ref{fig:hardstatebursts} and \citealt{PNK14}).
Soft state bursts of 4U 1608--52 do not show this behavior, nor do any other NS-LMXBs in the soft state \citep{KNL14}.
Our results indicate that the constant blackbody radii in the blackbody analysis may be caused by the spreading layer; the only way to obtain this is by having a constant colour correction factor as a function of flux, which is predicted by the spreading layer model \citep{SP06}.

The results provide important insights with respect to the NS mass and radius determination using PRE-bursts.
If the NS surface is partially covered by the spreading layer a number of assumptions mentioned in the introduction are not valid. 
Firstly, we do not see the whole undisturbed NS surface, but rather a part of the atmosphere is covered by the spreading layer, through which the burst emission has to penetrate.  
Secondly, a part (the lower half) of the NS surface is blocked by the optically thick accretion disc that is connected to the spreading layer.
Thirdly, the spreading layer has a colour correction factor that is in the range of $f_{\rm c}\sim 1.6$--$1.8$ irrespective of the flux.
When the soft-state bursts are used to determine the blackbody radii from the touchdown flux down to a small fraction of it \citep{GOC10,OPG16}, the radii are mapped into the NS radius and mass space by implicitly assuming that the touchdown flux is the Eddington flux, that the NS is entirely visible and that the colour correction factor lies in between $f_{\rm c} = 1.3$--$1.4$.
The latter two are clearly not correct assumptions, and probably not the first one either.
Note that the bursts shown in Fig. \ref{fig:softstatebursts}({a},{c}) were used by \citet{GOC10} to determine the Eddington flux $F_{\rm Edd}$ (more recently \citealt{OPG16} also added the burst shown in Fig. \ref{fig:softstatebursts2}d).
In these bursts, and in all bursts shown in Figs \ref{fig:softstatebursts} and \ref{fig:softstatebursts2} in general, one can see that the spreading layer component reprocesses roughly the half of the emitted burst flux.

There are still some uncertainties in the analysis that remain open.
For example, the ``bottom half'' of the NS should be covered by the accretion disc, but the touchdown fluxes are higher in the soft state compared to the hard state.
At first glance this observational fact could be considered to be inconsistent with our interpretation of the data.
However, the reflection of the burst emission off the inner accretion disc \citep{LS85} could cause stronger anisotropy of the emission in the soft state, where the inner disc edge is likely closer to the NS than in the hard state.
Thus, the burst emission could be boosted more in the soft-state bursts, as was argued by \citet{SPK16}.
However, in this case reflection features should be detected in the form of iron emission lines and photo-ionization edges.
Unlike in the superburst data where these features are clearly seen \citep{SB02, BS04, KBK14, KKK16}, the 4U 1608--52 soft state burst spectra do not require them.
Furthermore, when the burst emission irradiates the inner disc, a fraction of it may be absorbed rather than reflected enhancing the disc flux.
This effect does not seem to be present either.
As mentioned above, such disc heating should manifest as another NMF component, which is not detected in the analysis.
It is possible that we are not detecting these features in 4U 1608--52 simply because of the limited sensitivity compared to the superburst data, and thus future observations with \textit{Astrosat}, \textit{NICER} and/or \textit{eXTP} may help to address this issue. 

Another curious feature is related to the size scales during the first three seconds of the radius expansion phase, during which the \textsc{comptt} component carries a large fraction of the emitted flux. 
From the black body component we can estimate that black body radii are roughly 3 times larger during the PRE than in the burst tail. 
This indicates that the photospheric radius expansion is few tens of kilometers (note that here the unknown color correction factor in the PRE phase plays a critical role; see \citealt{PNK14,KNP17}). 
On the other hand, according to \citet{IS10} the radial width of the spreading layer is only a few hundred meters, and thus the layer (and the inner accretion disc) should be hidden underneath the expanded burst atmosphere.
The data clearly indicates that the spreading layer is visible, and it is not clear how this is possible. 
We can only speculate that the radius expansion is not symmetric for some reason, such that the expansion occurs predominately in the NS poles.

Another unclear issue is that there are some bursts where the spreading layer returns to its original state within 15~s (e.g. bursts in Fig. \ref{fig:softstatebursts}a and Fig. \ref{fig:softstatebursts2}a), whereas in others the spreading layer is active longer.
Also, some bursts are clearly much more non-Planckian than others and there does not appear to be any common factor behind these differences in terms of the burst onset conditions, i.e. the persistent emission level or the X-ray colours. 
The biggest issue, however, is clearly the difficulty to disentangle the two spectral components when the burst blackbody temperature is in the range of $\Tbb \sim 2.0$--$2.4$~keV.
Whenever the temperature passes through this range we see violent jumps from cold to hot phases in the model parameters during the PRE stage, and vice versa in the cooling tail.
The fact that such rapid changes -- and the corresponding \textsc{bb} component flux swings -- are not seen in the NMF decomposition means that they are without a doubt just modeling artifacts.
These discrepancies are likely related to the fitting differences.
In the \textsc{xspec} fitting, the \textsc{comptt} component is not allowed to vary in shape but only in normalization, whereas the blackbody can have any temperature or normalization.
Because in the $\Tbb \sim 2.0$--$2.4$~keV range the blackbody and the \textsc{comptt} produce very similar spectral shapes, the \textsc{comptt} remains a stable component as only its flux is allowed to vary, and the two blackbody parameters then act to compensate the small residuals left in the data. 
In contrast in the NMF analysis all the burst spectra are decomposed into three linear components that all have fixed ``shapes'', and therefore these components do not compete in the cost function minimization when the summed spectral shape of the $k=2,3$ components corresponds to the $k=1$ shape.

Therefore, the most likely scenario is that a small polar cap is always directly visible, with the rest of the NS being covered underneath the spreading layer at touchdown. 
Once the flux starts decaying and the atmosphere cools, the NS also becomes gradually more visible.
We speculate that -- as the NS is oblate given its large spin frequency of $\approx 620$~Hz \citep{Watts2012} -- the pole should have larger gravitational pull than the equatorial regions, and thus the burst blackbody can attain higher Eddington temperatures at the poles compared to the regions covered by the spreading layer.
However, we cannot completely rule out a possibility that the entire NS is covered by the spreading layer until the point in time when the temperature jump occurs, and the hottest parts of the bursts may just be caused by the spreading layer spectrum changing its temperature when the entire NS becomes covered. 
For example, in the burst shown in Fig.~\ref{fig:softstatebursts2}(d) the cooling seems to stall for up to five seconds, during which the spectrum is identical to the spreading layer.
This similarity favours the latter interpretation, but the fact that the peak temperatures are $\Tbb \sim 3.0$~keV, just like in the hard-state bursts, lends support to the former scenario. 
From the 4U 1608--52 data alone it is not clear which one actually happens and a more systematic comparison between various bursters is warranted in the future.
It is also not clear how scalable these results are with respect to other bursters.
While some sources show similar non-Planckian spectra as 4U 1608--52 in the soft state and correspondingly constant blackbody radii in the initial cooling tail \citep{GOP12}, other sources do not.
Here, however, it is important to study the bursts as a function of spectral state (as in \citealt{ZMA11}), as in the hard state other effects than the spreading layer may also distort the persistent spectra, for example,  the burst-induced cooling of the coronae \citep{JZC15, DKC16, KSK17}, or by the intermittent presence of  metals in the NS photospheres \citep{NSK15,KNP17}.

\section{Summary}

In this paper we have presented a study of X-ray bursts from the LMXB 4U 1608--52, mainly focusing on the PRE bursts that occur while the system is in the soft spectral state.
Using the non-parametric NMF spectral decomposition method, we found that during most of the non-Planckian PRE bursts in the soft state there are in fact two variable spectral components during the bursts.
One component is well described with a variable temperature X-ray burst black body component, while the other component is identical to the spreading layer that forms in the interface between the accretion disc and the NS surface.
We find that during the bursts the spreading layer component brightens by a factor up to about 50.
This occurs, in our view, because the enhanced radiation pressure provided by the X-ray burst emission underneath the spreading layer pushes it towards the NS poles, possibly engulfing the entire NS surface during the brightest phases of the bursts.

This physical picture provides an alternative interpretation to the apparent increase of the persistent emission level during the bursts through the Poynting-Robertson drag mechanism, as proposed by \citet{WGP13,WGP15}.
Instead, we argue that the burst emission is reprocessed in the spreading layer as it passes through it, causing the layer to have a variable latitudinal width in the course of the bursts, which only mimics an increase of the persistent emission.

Due to this interplay, the burst emission gets distorted and thus the spectral color correction factor, that is used in determining NS masses and radii is much higher than for a passively cooling NS atmosphere.
Furthermore, as the layer is optically thick, a fraction of the NS that would be visible due to light bending effects is hidden in these soft state bursts.
These two factors together are the likely causes for the significantly different NS mass and radius constraints obtained using the soft state bursts vs. the hard state bursts, where the spreading layer does not play a role in shaping the burst spectra.

\section*{Acknowledgements}

We thank the anonymous referee for helpful suggestions that improved the manuscript.
JJEK was supported by Academy of Finland grants 268740 and 295114.
JN was supported by the University of Turku Graduate School in Physical and Chemical Sciences.
VFS was supported by the German Research Foundation (DFG) grant WE 1312/48-1 and by the Russian Government Program of Competitive Growth of Kazan Federal University. 
JP thanks  the Foundations' Professor Pool, the Finnish Cultural Foundation and the National Science Foundation grant PHY-1125915 for support.


\bibliographystyle{mnras}

\bsp	
\label{lastpage}
\end{document}